\documentclass[12pt,letterpaper]{article}
\pdfoutput=1
\usepackage{jheppub}
\usepackage{dsfont}
\usepackage{amsmath,amssymb,amscd,amsfonts,mathtools}

\DeclareMathOperator{\Tr}{Tr}

\usepackage[toc,page]{appendix}
\usepackage{epsfig}
\usepackage{epstopdf}
\usepackage{latexsym}
\usepackage{graphicx}
\usepackage{subfig}
\usepackage{booktabs}
\usepackage{bbm}
\usepackage{color}
\usepackage{physics}
\usepackage{tensor}
\usepackage{hyperref}

\usepackage{tikz}
\usetikzlibrary{matrix}
\usetikzlibrary{decorations.markings,calc,shapes,decorations.pathmorphing,patterns,decorations.pathreplacing}
\usetikzlibrary{positioning}

\numberwithin{equation}{section}
\allowdisplaybreaks

\makeatletter
\def\@fpheader{\relax}
\makeatother

\subheader{\begin{flushright}
UTTG-06-21
\end{flushright}}

\title{Circuit Complexity in Topological Quantum Field Theory}
\author{Josiah Couch,$^{a, b}$ Yale Fan,$^a$ Sanjit Shashi$^a$}
\affiliation{$^a$Theory Group, Department of Physics, University of Texas, Austin, TX 78712, USA}
\affiliation{$^b$Department of Computer Science, Boston College, Chestnut Hill, MA 02467, USA}

\emailAdd{couchjo@bc.edu} % josiah.couch@utexas.edu
\emailAdd{yalefan@gmail.com}
\emailAdd{sshashi@utexas.edu}

\abstract{Quantum circuit complexity has played a central role in recent advances in holography and many-body physics. Within quantum field theory, it has typically been studied in a Lorentzian (real-time) framework. In a departure from standard treatments, we aim to quantify the complexity of the Euclidean path integral. In this setting, there is no clear separation between space and time, and the notion of unitary evolution on a fixed Hilbert space no longer applies. As a proof of concept, we argue that the pants decomposition provides a natural notion of circuit complexity within the category of 2-dimensional bordisms and use it to formulate the circuit complexity of states and operators in 2-dimensional topological quantum field theory. We comment on analogies between our formalism and others in quantum mechanics, such as tensor networks and second quantization.}

\begin{document}
\setcounter{tocdepth}{3} % 2 or 3
\maketitle
\flushbottom

%%%%%%%%%%%%%%%%%
\section{Introduction}\label{sec:intro}
%%%%%%%%%%%%%%%%%

The past decade has seen considerable interest in the relevance of quantum circuit complexity\footnote{Quantum circuit complexity is roughly the number of basic quantum operations, or \emph{gates}, needed to construct a quantum operator or to prepare a quantum state from a simple reference state.} to holography and quantum gravity. Following Harlow and Hayden's proposed complexity-theoretic resolution of the firewall paradox \cite{Harlow:2013tf}, this interest gained traction with the proposal of potential holographic duals to the complexity of CFT states \cite{Susskind:2014rva, Stanford:2014jda, Susskind:2014moa, Brown:2015bva, Brown:2015lvg}. These developments have in turn motivated various attempts to understand complexity in the context of quantum field theory \cite{Hashimoto:2017fga, Jefferson:2017sdb, Chapman:2017rqy, Khan:2018rzm, Hackl:2018ptj, Guo:2018kzl, Bhattacharyya:2018bbv, Chapman:2018hou, Caceres:2019pgf, Magan:2018nmu, Caputa:2018kdj, Bueno:2019ajd, Erdmenger:2020sup, Flory:2020eot, Flory:2020dja, Chagnet:2021uvi, Koch:2021tvp, Camilo:2019bbl, Leigh:2021trp, Meng:2021wmz, Moghimnejad:2021rqe}, a markedly different setting than that of the many-qubit systems in which complexity is usually discussed.

This paper represents another such attempt, albeit one that takes a rather different approach to the problem. Rather than invoking a lattice regularization and reducing the problem of complexity to one in quantum mechanics---or, indeed, referring to any complexity measure on a space of unitaries at all---we take as our starting point the axioms of quantum field theory.

It has long been known that a quantum field theory, depending on its assumed symmetries, can be defined as a \emph{functor} from an appropriate category of bordisms to the category of Hilbert spaces.\footnote{A category is a mathematical structure that consists of objects (e.g., sets, groups, vector spaces) along with ``arrows'' or maps between them (e.g., functions, homomorphisms, linear operators). A functor is a homomorphism of categories. See Appendix \ref{app:category} for details.} A bordism between $(d - 1)$-dimensional closed manifolds $A$ and $B$ is a $d$-dimensional compact manifold whose boundary is the disjoint union of $A$ and $B$. In $d$ spacetime dimensions, the QFT functor (understood as evaluation of the path integral) identifies $(d - 1)$-dimensional closed manifolds with Hilbert spaces and $d$-dimensional bordisms with linear operators between Hilbert spaces in such a way that gluing of bordisms goes to composition of linear maps. Such an axiomatization is most well-established for topological and conformal field theories \cite{Atiyah:1989vu, Segal1988, Runkel:2005qw, henriques2013threetier}.

In this paper, we propose that a quantum field theory inherits a notion of complexity for states and operators from one on its corresponding category of bordisms. The latter form of complexity is determined by the geometry of the \emph{Euclidean path integral}.\footnote{Our perspective, while independent of holographic considerations, bears some conceptual similarities to tensor network toy models of holography \cite{Vidal:2008zz, Swingle:2009bg, Evenbly_2015, Miyaji:2016mxg, Caputa:2017urj, Caputa:2017yrh, Camargo:2019isp, Bhattacharyya:2019kvj, Chandra:2021kdv}, in which the geometry of a spatial slice is understood as a quantum circuit and thereby provides a natural notion of complexity for boundary states.}

Our concrete implementation of this idea takes place in the setting of \textit{topological} quantum field theory (TQFT), the simplest class of QFT that can be described in functorial language. A $d$-dimensional TQFT is a functor from the category of $d$-dimensional bordisms up to homeomorphism (called $d\textbf{Cob}$) to the category of Hilbert spaces \textbf{Hilb}:
\begin{equation}
Z: d\textbf{Cob} \to \textbf{Hilb}.
\end{equation}
Topological field theories circumvent the complications of infinite-dimensional Hilbert spaces, and their axiomatization renders the path integral completely finite and unambiguous.

Our task is thus to formulate a definition of computational complexity in functorial TQFT. To simplify this task, it is useful to narrow our attention to certain TQFTs for which the basic principles of complexity can be made especially clear. In dimension $d = 1$, the limited structure of bordisms allows for no possibility of preparing a nontrivial manifold of target states from any reference state (1D TQFTs are simply isomorphic to finite-dimensional vector spaces). On the other hand, in dimensions $d\geq 3$, the objects of $d\textbf{Cob}$ are not finitely generated with respect to disjoint union, so a $d$-dimensional TQFT requires an infinite amount of algebraic data to specify \cite{Carqueville:2017fmn}. We are therefore led to focus on the tractable and still interesting case of 2D TQFT \cite{Peelaers:2020ncm, Kock:2004cob}. In $d = 2$, bordisms allow for nontrivial propagation in spacetime, yet the category $2\textbf{Cob}$ admits a presentation in terms of finitely many generators and relations.

In fact, the category $2\textbf{Cob}$ comes equipped with a natural notion of complexity. Roughly speaking, any $2$-dimensional bordism can be decomposed into disks and pairs of pants, which provide a finite and exactly universal\footnote{The set of elementary gates inherent in the definition of complexity should be \emph{universal} in the sense of enabling the construction of all gates or states of interest, whether exactly or to any desired accuracy.} ``gate set'' on $2\textbf{Cob}$. The complexity of a given bordism follows from counting the minimum number of these component surfaces needed to construct it.

Given that a 2D TQFT can be thought of as a functor from this category to that of Hilbert spaces, our main claim is that complexity in $2\textbf{Cob}$ \textit{induces} a useful notion of complexity on the image of the TQFT functor in \textbf{Hilb}. We interpret the latter as a quantum complexity in 2D TQFT.

The remainder of this paper is devoted to expounding this idea. It is structured as follows. In Section \ref{sec:complexity}, we start by reviewing various notions of quantum complexity and adapting them to our context. In Section \ref{sec:functorial}, we assemble the tools of functorial TQFT that feature in our analysis, focusing on the classification of 2D TQFTs and their equivalence to commutative Frobenius algebras. In Section \ref{sec:inducedComplexity}, we develop the idea of induced complexity in functorial TQFT. A key point is that the image of the natural gate set of $2\textbf{Cob}$ under the TQFT functor does not generically allow for the construction of arbitrary states. Indeed, one might (correctly) suspect that due to the mismatch between the discrete infinity of topologies and the continuous infinity of states, the TQFT path integral has no hope of populating the entire TQFT Hilbert space. We instead focus on the class of \emph{semisimple} 2D TQFTs, for which the path integral allows for the construction of a continuous submanifold of states (which turns out to be a torus) within the Hilbert space associated to an arbitrary spatial manifold. In Section \ref{sec:applicationsExtensions}, we discuss interesting extensions of this point of view.

Two appendices contain some useful background material. In Appendix \ref{app:category}, we provide a self-contained review of the relevant elements of category theory. In Appendix \ref{app:frobenius}, we summarize the classification of ``two-level'' 2D TQFTs (2D TQFTs with 2D Hilbert space on a circle), which serve as examples throughout this paper.

%%%%%%%%%%%%%%%%%%%%%%%%%%
\section{Complexity} \label{sec:complexity}
%%%%%%%%%%%%%%%%%%%%%%%%%%

Before studying complexity in specific physical systems, we establish some key concepts by discussing complexity in groups and in categories. The idea of complexity in these two mathematical settings closely mirrors that in quantum mechanics and quantum field theory, respectively. In all of these contexts, we distinguish between two types of circuit complexity: operator complexity and state complexity.

%%%%%%%%%%%%%%%%%%%%%%%%%%
\subsection{Groups} \label{subsec:groups}
%%%%%%%%%%%%%%%%%%%%%%%%%%

Let $G$ be a group. Any generating set $S\subset G$ (possibly infinite) can be said to form a \textit{universal gate set} on $G$ in the sense that any element $g \in G$ admits a decomposition into a finite product or \textit{circuit} $g = s_1\cdots s_k$, where $s_i\in S$. We define the \textit{size} of such a circuit to be the number of generators $k$ appearing therein, and we define the \textit{circuit complexity} $\mathcal{C}(g)$ to be the smallest size of any circuit that evaluates to $g$. We further define the complexity of $g$ relative to a reference group element $g_0$ as $\mathcal{C}(g, g_0)\equiv \mathcal{C}(gg_0^{-1})$. Assuming that $S$ is closed under inverse ($S = S^{-1}$), the relative complexity $\mathcal{C}(g, h)$ for $g, h\in G$ defines a metric on $G$---the ``word metric'' with respect to $S$ \cite{Lin:2018cbk}.

In practice, we wish to restrict to finite $S$, which leads us to distinguish the notions of exact and approximate universality. If $G$ is finitely generated, then every $g\in G$ can be prepared exactly by a finite-size circuit drawn from a finite $S$. Otherwise, we call a finite gate set $S\subset G$ \emph{universal up to the tolerance $\epsilon$} if any $g\in G$ can be prepared to within a given tolerance $\epsilon > 0$ by a circuit constructed from $s_i\in S$. More precisely, we equip $G$ with a metric $d(\cdot, \cdot)$ and say that a circuit of size $k$ prepares $g$ if $d(g, s_1\cdots s_k)\leq \epsilon$. We denote the corresponding circuit complexity by $\mathcal{C}_\epsilon(g)$.

Now let $X$ be a set equipped with a transitive group action by $G$. We can define the \textit{state complexity} of any $x\in X$ relative to some reference $x_0 \in X$ as the minimum circuit complexity of any group element $g$ for which $x = gx_0$. In other words, operator complexity in $G$ induces state complexity in $X$. If the group action is not transitive, then only elements in the orbit of $x_0$ have a sensible complexity. This definition can be extended to accommodate approximate universality on both $G$ and $X$.

%%%%%%%%%%%%%%%%%%%%%%%%%%
\subsection{Categories} \label{subsec:categories}
%%%%%%%%%%%%%%%%%%%%%%%%%%

Categories are structures with both objects and morphisms. Operator complexity in this case is tantamount to ``morphism complexity,'' which is analogous to circuit complexity on a group. Setting aside issues of universality and tolerance, it is defined as the minimum number of appropriately chosen ``elementary morphisms'' needed to construct a given morphism by composition. On the other hand, state complexity assigns a complexity to every state, given a reference state. This is analogous to complexity on a space that carries a group action.

To make sense of the latter notion, we must first define a ``state'' in the categorical setting. This is easy to do for a \emph{monoidal} category (see Appendix \ref{app:category}), which in particular has a distinguished unit object. Given any object $x$, we may define a state as a morphism $1 \to x$ where $1$ is the unit object. If we fix a reference state $r : 1 \to x$, then we can obtain another state $f : 1\to y$ by composing with a morphism $\mathcal{O} : x\to y$, and we can identify the state complexity of $f$ as the minimum operator complexity over all $\mathcal{O}$'s (as morphisms) such that $\mathcal{O} \circ r = f$. By choosing $x = y$, we restrict to only those states ``belonging to'' the object $x$.

Note that state complexity, as we have defined it, is not synonymous with ``object complexity'' in the sense of the minimum number of elementary morphisms needed to go between a reference object $x$ and a target object $y$. Object complexity is coarser than state complexity because every object can be associated with a multitude of states. However, object and state complexity would coincide if we were to instead define a state as a morphism $I\to x$ where $I$ is an \emph{initial} object of the category.\footnote{An initial object $I$ in a category is an object such that for every object $x$, there is precisely one morphism $I\to x$.}

While the above definitions may seem abstract, their motivation is decidedly non-esoteric. Consider the example of classical computation. A classical circuit implements a function from $n$ bits to $m$ bits, and its circuit complexity is defined with respect to a universal gate set (e.g., the singleton set $\{\text{NAND}\}$). Not only are such functions generally not invertible, but they generally do not even have the same domain and codomain. The case of classical circuit complexity makes clear that complexity is not a notion restricted to groups.\footnote{\emph{Reversible} classical computation, however, corresponds to the case of the permutation group on $2^n$ elements acting on the set of all $n$-bit strings \cite{Lin:2018cbk}.}

Rather, we may regard classical circuit complexity as being defined on a category where each object is the set of all $n$-bit strings $\{0, 1\}^n$ for some $n\geq 0$, and where the morphisms are set functions. This category has a symmetric monoidal structure with respect to Cartesian product (element-wise concatenation), where the unit object is the set consisting of the empty string. We may then take certain of these set functions to be gates, circuits to be compositions of gates, and the complexity of an arbitrary function to be the minimum number of gates which must be composed to build that function. Note that a logic gate in the conventional sense (e.g., a Boolean function $\{0, 1\}^2\to \{0, 1\}$ or $\{0, 1\}\to \{0, 1\}$) actually gives rise to an infinite family of gates in the categorical sense, each coming from starting with a different number of bits and then choosing a particular subset of those bits as input to the logic gate.

Returning to physics, our primary interest will lie in the monoidal categories 2\textbf{Cob} (where the unit object with respect to disjoint union is the empty set) and \textbf{Hilb} (where the unit object with respect to tensor product is $\mathbb{C}$). Our definition of state complexity makes as much sense for 2\textbf{Cob} as for \textbf{Hilb}, but is richer in \textbf{Hilb} due to the far greater variety of reference states and target states.

%%%%%%%%%%%%%%%%%%%%%%%%%%%%%%%%%%%%%%%
\subsection{Quantum Mechanics} \label{subsec:qmech}
%%%%%%%%%%%%%%%%%%%%%%%%%%%%%%%%%%%%%%%

The usual notion of quantum complexity corresponds to the case of the unitary group $U(\mathcal{H})$ acting on a Hilbert space $\mathcal{H}$.\footnote{It is physically more appropriate to say that the action is on the projective Hilbert space $\mathbb{P}(\mathcal{H})$. When we discuss TQFT, we will switch between $\mathcal{H}$ and $\mathbb{P}(\mathcal{H})$, but note that physics only distinguishes between states in the latter.} There are thus two distinct formulations of complexity in quantum computing: circuit complexity for unitary transformations and state complexity for quantum states \cite{Aaronson:2016vto}. Since $U(\mathcal{H})$ is not finitely generated, a tolerance $\epsilon$ is assumed:
\begin{itemize}
\item The complexity of an operator $O$ is the smallest number of gates in any circuit $Q$ such that $|O-Q|\leq \epsilon$, where we can take $|\cdot|$ to be the operator norm.
\item The complexity of a state $\ket{\psi}$ relative to a reference state $\ket{r}$ is the minimum complexity of any unitary $U$ such that $U\ket{r} = \ket{\psi}$.
\end{itemize}
The definition of quantum circuit complexity suffers from a number of ambiguities involving the choice of universal quantum gate set, the choice of tolerance parameter, and (in the case of state complexity) the choice of reference state.

Again, note that the word ``gate'' as used in the definitions above does not quite correspond to a gate in the conventional sense. For instance, the CNOT gate of quantum computing does not correspond to a single operator on the Hilbert space of $n$ qubits, but rather to $n(n - 1)$ different operators, obtained by applying CNOT to specific qubits in all possible ways. In the above definitions, each of these $n(n - 1)$ operators must be included as a separate gate.

The notion of complexity that we will employ in this work is discrete, in contrast to the continuous ``complexity geometry'' approach that relies on a metric on the space of unitary operators on the relevant Hilbert space \cite{Nielsen:2005cg, Nielsen_20061, Nielsen:2006cg}. Hence ``universality'' will always refer to the constructibility of unitaries or the ability to prepare arbitrary states to a given accuracy.

%%%%%%%%%%%%%%%%%%%%%%%%%%%%%%%%%%%%%%%%%
\subsection{Quantum Field Theory} \label{subsec:qft}
%%%%%%%%%%%%%%%%%%%%%%%%%%%%%%%%%%%%%%%%%

It is not immediately clear how to extend complexity to a continuum field theory. For instance, what should we take as the reference state in a field theory setting, and how should we choose a set of elementary gates to ensure universality? The interest in holographic complexity has spurred much initial work on answering these questions: definitions of circuit complexity in QFT have been proposed for abelian gauge theory \cite{Hashimoto:2017fga, Meng:2021wmz, Moghimnejad:2021rqe}, free scalar field theory \cite{Jefferson:2017sdb, Chapman:2017rqy, Guo:2018kzl, Chapman:2018hou, Caceres:2019pgf}, free fermionic field theory \cite{Khan:2018rzm, Hackl:2018ptj}, interacting scalar field theory \cite{Bhattacharyya:2018bbv}, 2D CFT \cite{Magan:2018nmu, Caputa:2018kdj, Bueno:2019ajd, Erdmenger:2020sup, Flory:2020eot, Flory:2020dja}, higher-dimensional CFT \cite{Chagnet:2021uvi, Koch:2021tvp}, and Chern-Simons theory \cite{Camilo:2019bbl, Leigh:2021trp}. In general, however, these efforts have not identified a universal gate set in their particular settings. For instance, discussions of free field theory and 2D CFT tend to focus on non-universal gate sets (e.g., ``symmetry gates'') that relate only Gaussian states and states in the same Verma module, respectively.

More broadly, previous treatments of circuit complexity in QFT have focused on unitary time evolution on a fixed spatial geometry.\footnote{Notable exceptions include the ``path integral optimization'' of \cite{Miyaji:2016mxg, Caputa:2017urj, Caputa:2017yrh} as well as the definition of relative complexity for holographic CFT states proposed in \cite{Belin:2018fxe, Belin:2018bpg}. The latter is based on the distance between coherent states prepared by the Euclidean path integral with sources for single-trace operators. These Euclidean methods rely on continuous cost or distance functionals, unlike ours.} One crucial point missing from these discussions is that while QFT can be formulated in Lorentzian signature on spacetime manifolds of the form ``space $\times$ time,'' it can also be formulated in Euclidean signature on spacetimes of arbitrary topology. This leads to possibilities beyond the Hamiltonian evolution familiar from quantum mechanics, and is responsible for much of the richness of QFT. In this paper, we address questions of state preparation and complexity in the context of the Euclidean path integral. Our central point is that topology change in the Euclidean path integral can have nontrivial computational effects, particularly via the implementation of inherently non-unitary transformations. These effects are intrinsic to QFT and absent from quantum mechanics.

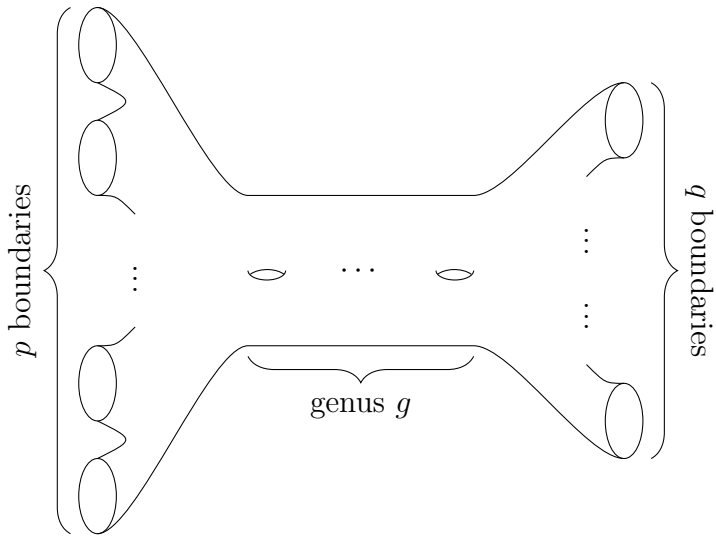
\begin{figure}
\centering
\begin{tikzpicture}
\draw[-] (-2,3) ellipse (0.25 and 0.5);
\draw[-] (-2,1.5) ellipse (0.25 and 0.5);
\node at (-1.5,0) {$\vdots$};
\draw[-] (-2,-3) ellipse (0.25 and 0.5);
\draw[-] (-2,-1.5) ellipse (0.25 and 0.5);

\draw[-] (-2,2.5) .. controls (-1.5,2.25) .. (-2,2);
\draw[-] (-2,-2.5) .. controls (-1.5,-2.25) .. (-2,-2);

\draw[-] (-2,1) .. controls (-1.75,1) .. (-1.5,0.75);
\draw[-] (-2,-1) .. controls (-1.75,-1) .. (-1.5,-0.75);

\draw[-] (-2,3.5) .. controls (-1.5,3.5) and (-0.5,1) .. (0,1) to (3,1) .. controls (3.5,1) and (4.5,2.5) .. (5,2.5);
\draw[-] (-2,-3.5) .. controls (-1.5,-3.5) and (-0.5,-1) .. (0,-1) to (3,-1) .. controls (3.5,-1) and (4.5,-2.5) .. (5,-2.5);

\draw[-] (5,2) ellipse (0.25 and 0.5);
\draw[-] (5,-2) ellipse (0.25 and 0.5);

\draw[-] (5,1.5) .. controls (4.75,1.5) .. (4.5,1.25);
\draw[-] (5,-1.5) .. controls (4.75,-1.5) .. (4.5,-1.25);

\node at (4.5,0.5) {$\vdots$};
\node at (4.5,-0.5) {$\vdots$};

\draw [decorate,decoration={brace,amplitude=10pt},xshift=-4pt,yshift=0pt]
(5.5,2.5) -- (5.5,-2.5);
\node[rotate=-90] at (6,0) {$q$ boundaries};

\draw [decorate,decoration={brace,amplitude=10pt},xshift=4pt,yshift=0pt]
(-2.5,-3.5) -- (-2.5,3.5);
\node[rotate=90] at (-3,0) {$p$ boundaries};

\draw[-] (0,0) arc (180:360:0.25 and 0.125);
\draw[-] (0.03,-0.05) arc (145:35:0.275 and 0.15);

\draw[-] (3,0) arc (0:-180:0.25 and 0.125);
\draw[-] (3-0.03,-0.05) arc (35:145:0.275 and 0.15);

\node at (1.5,0) {$\cdots$};

\draw [decorate,decoration={brace,amplitude=10pt},yshift=-4pt,xshift=0pt]
(3,-1) -- (0,-1);

\node at (1.5,-1.825) {genus $g$};

\end{tikzpicture}
\caption{A bordism representing a map $\mathcal{H}_{S^1}^{\otimes p} \to \mathcal{H}_{S^1}^{\otimes q}$ in a 2D TQFT.}
\label{fig:cobMap}
\end{figure}

This point is especially stark in the context of TQFT, where the \emph{only} nontrivial amplitudes arise from topology change. Our definition of complexity in 2D TQFT, to be made more precise later on, goes roughly as follows. Let $\mathcal{H}_{S^1}$ denote the Hilbert space on $S^1$. Consider states $\ket{\psi_p} \in \mathcal{H}_{S^1}^{\otimes p}$ and $\ket{\psi_q} \in \mathcal{H}_{S^1}^{\otimes q}$, prepared by path integrals on oriented surfaces with $p$ and $q$ outward-oriented circle boundaries, respectively. Consider further an oriented bordism from $p$ ``in-circles'' to $q$ ``out-circles'' composed of some network of tubes and handles, as shown in Figure \ref{fig:cobMap} (we also allow for the possibility of multiple connected components). Finally, consider composing this bordism with $\ket{\psi_p}$ to obtain a state $\ket{\psi_q'}\in \mathcal{H}_{S^1}^{\otimes q}$. Then topological circuit complexity should quantify the simplest bordism needed to approximate $\ket{\psi_q}$ by $\ket{\psi_q'}$, starting from a given $\ket{\psi_p}$. If our topological ``gate set'' is non-universal, then it may not be possible to achieve this approximation to arbitrary accuracy.

It is convenient to consider bordisms with only out-boundaries ($p = 0$). This setup allows for various choices of reference state. For example, one could take as a reference the state corresponding to $q$ copies of the disk, each viewed as a bordism from $\varnothing$ to $S^1$. An arbitrary state, viewed as a bordism from $\varnothing$ to $(S^1)^q$, could then be constructed as the composition of a bordism from $(S^1)^q$ to $(S^1)^q$ with this reference state. This perspective has the advantage of operating within a fixed Hilbert space, namely $\mathcal{H}_{S^1}^{\otimes q}$.

However, we emphasize that the choice of reference state is not essential to our complexity-theoretic considerations. Physically, it makes sense to consider arbitrary reference states because any state in the TQFT Hilbert space can be evolved according to the TQFT dynamics, even if the preparation of the state is external to the TQFT. Indeed, most states lie outside the image of the TQFT functor and thus cannot be represented in path integral language as bordisms from the empty set.

Our definition of complexity in TQFT differs fundamentally from the usual definition of quantum circuit complexity (which concerns unitary transformations between states in a fixed Hilbert space) in that the linear maps associated to bordisms are typically not invertible or unitary.\footnote{Early investigations of the computational power of TQFT \cite{Freedman:2000rc, Freedman2000AMF} were also restricted to unitary transformations on a fixed Hilbert space, or the strictly ``2D part'' of a unitary 3D TQFT (formalized as a \emph{unitary topological modular functor}). We, by contrast, consider arbitrary bordisms.} The irreversible nature of topological gates resembles that of classical logic gates. This is a manifestation of the fact that, unlike in quantum mechanics, the Hilbert space is not fixed at all intermediate (Euclidean) times. Since time evolution along the cylinder is trivial (the Hamiltonian vanishes in a TQFT), our construction quantifies precisely those aspects of complexity that are intrinsic to topology change.

%%%%%%%%%%%%%%%%%
\section{Functorial TQFT} \label{sec:functorial}
%%%%%%%%%%%%%%%%%

We now introduce the functorial language in which the ideas of the previous section can be made precise. We provide only the barest introduction to the formalism, leaving additional details to Appendix \ref{app:category}. Throughout, we work over $\mathbb{C}$, and we assume smooth manifolds with boundary.

%%%%%%%%%%%%%%%%%%%%%%%%%%%%%%%%%
\subsection{Generalities} \label{subsec:general}
%%%%%%%%%%%%%%%%%%%%%%%%%%%%%%%%%%

Following Atiyah \cite{Atiyah:1989vu}, a $d$-dimensional TQFT $Z$ is a ``rule'' that assigns a vector space $Z(\Sigma)$ (state space) to every closed, oriented $(d-1)$-dimensional manifold $\Sigma^{d-1}$ (spatial slice) and a linear map $Z(M) : Z(\Sigma_\text{in})\to Z(\Sigma_\text{out})$ (time evolution operator) to every oriented $d$-dimensional bordism $M^d$ with boundary $\partial M = \Sigma_\text{in}\sqcup \Sigma_\text{out}$ (spacetime). This rule satisfies several physically motivated axioms, as encapsulated by the following definition:
\begin{quote}
\textit{A $d$-dimensional TQFT is a symmetric monoidal functor $Z$ from the category} $d$\textbf{Cob} \textit{of $d$-dimensional bordisms to the category} \textbf{Vect}$_{\mathbb{C}}$ \textit{of vector spaces over $\mathbb{C}$, or equivalently, a linear representation of} $d$\textbf{Cob}.
\end{quote}
(This definition makes no reference to an inner product, so we phrase it in terms of \textbf{Vect}$_{\mathbb{C}}$ rather than the category of Hilbert spaces \textbf{Hilb}.)

In physical terms, the path integral on a closed manifold computes a linear map $\mathbb{C}\to \mathbb{C}$ (the partition function), while the path integral on a manifold with outward-oriented boundary prepares a boundary state (interpreted as a linear map from $\mathcal{H}(\varnothing) = \mathbb{C}$ to the boundary Hilbert space). Such boundary states are wavefunctionals of boundary field configurations---fixing boundary conditions for the fields yields the state evaluated at a particular point in field space. A choice of boundary conditions is a choice of basis for the boundary Hilbert space.

As a special case, a \textit{unitary} TQFT is a symmetric monoidal functor $Z$ from $d$\textbf{Cob} to \textbf{Hilb} that satisfies $Z(\overline{M}) = Z(M)^\dag$, where bar denotes orientation reversal and dagger denotes adjoint \cite{Durhuus:1993cq, Sawin:1995rh}. Note that the axiomatic definition of a ``unitary'' (or more precisely, reflection-positive) TQFT is broader than the Hamiltonian definition of unitarity in Lorentzian signature. Indeed, axiomatic TQFTs can be thought of as formulated in Euclidean signature (imaginary time), where the linear operators associated to bordisms are not generally unitary and therefore act on projective Hilbert space. The Lorentzian notion of unitarity applies in the absence of topology change.

It will be important for us that there exists a direct sum construction of TQFTs \cite{Durhuus:1993cq}. Given two $d$-dimensional TQFTs $Z'$ and $Z''$, their direct sum $Z = Z'\oplus Z''$ associates to any connected $\Sigma$ the vector space $Z(\Sigma)\equiv Z'(\Sigma)\oplus Z''(\Sigma)$, and to any connected $M$ the linear map $Z(M)\equiv Z'(M)\oplus Z''(M)$. The map extends to disconnected $(\Sigma, M)$ via tensor products.

%%%%%%%%%%%%%%%%%
\subsection{2D TQFT}\label{subsec:2dTQFT}
%%%%%%%%%%%%%%%%%

A 2-dimensional (2D) TQFT is a symmetric monoidal functor from 2\textbf{Cob} to \textbf{Vect}$_{\mathbb{C}}$. 2\textbf{Cob} is the monoidal category of oriented bordisms in 2D. Its objects are closed, oriented 1-manifolds (disjoint unions of circles). A morphism from $\Sigma$ to $\Sigma'$ is an oriented 2-manifold whose in-boundary is $\Sigma$ and whose out-boundary is $\Sigma'$ (depending on whether the positive normal vector points inward or outward). Two bordisms are regarded as equivalent ($\cong$) if they are related by an orientation-preserving diffeomorphism relative to the boundary.\footnote{Since two smooth surfaces are diffeomorphic if and only if they are homeomorphic, we use homeomorphism from now on; moreover, every topological surface admits a smooth structure.}

%%%%%%%%%%%%%%%%%%%%%%%%%%%%%%%%%%%%%%%%%%%
\subsubsection{Frobenius Algebras} \label{subsubsec:frobenius}
%%%%%%%%%%%%%%%%%%%%%%%%%%%%%%%%%%%%%%%%%%%

It is well-known that every 2D TQFT is uniquely specified by a \emph{commutative Frobenius algebra}. In particular, a unitary 2D TQFT is a commutative $H^\ast$-algebra, also called a $C^\ast$-Frobenius algebra \cite{Sawin:1995rh}. (See \cite{Kock:2004cob} for a review and historical account of this equivalence, and \cite{Carqueville:2017fmn, Peelaers:2020ncm} for modern reviews of low-dimensional TQFT.)

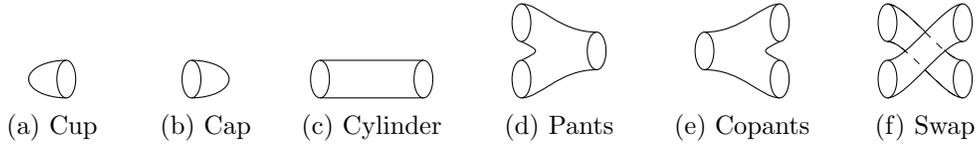
\begin{figure}
\centering
\newcommand{\globalscale}{0.5} % 0.65

\subfloat[Cup\label{figs:cup}]
{
\begin{tikzpicture}[rotate=90,scale=\globalscale]
\node at (0,-0.65) {};
\node at (0,1.4) {};
\draw[-] (-0.5,0) arc (180:0:0.5 and 0.25);
\draw[-] (-0.5,0) arc (180:360:0.5 and 0.25);
\draw[-] (-0.5,0) arc (180:0:0.5 and 1);
\end{tikzpicture}
}\qquad\hspace{-0.5cm}
\subfloat[Cap\label{figs:cap}]
{
\begin{tikzpicture}[rotate=-90,scale=\globalscale]
\node at (0,-0.65) {};
\node at (0,1.4) {};
\draw[-] (-0.5,0) arc (180:0:0.5 and 0.25);
\draw[-] (-0.5,0) arc (180:360:0.5 and 0.25);
\draw[-] (-0.5,0) arc (180:0:0.5 and 1);
\end{tikzpicture}
}\qquad\hspace{-0.5cm}
\subfloat[Cylinder\label{figs:cyl}]
{
\begin{tikzpicture}[scale=\globalscale]
\draw[-] (-0.5,0) arc (180:0:0.25 and 0.5);
\draw[-] (-0.5,0) arc (180:360:0.25 and 0.5);

\draw[-] (2.25,0) arc (180:0:0.25 and 0.5);
\draw[-] (2.25,0) arc (180:360:0.25 and 0.5);

\draw[-] (-0.25,0.5) to (2.5,0.5);
\draw[-] (-0.25,-0.5) to (2.5,-0.5);
\end{tikzpicture}
}\quad\quad\hspace{-0.2cm}
\subfloat[Pants\label{figs:pants}]
{
\begin{tikzpicture}[xscale=-1,scale=\globalscale]
\draw[-] (0,0) ellipse (0.25 and 0.5);

\draw[-] (0,0.5) .. controls (1,0.5) and (1.5,1.25) .. (2,1.25);
\draw[-] (2,0.25) .. controls (1.5,0) .. (2,-0.25);
\draw[-] (0,-0.5) .. controls (1,-0.5) and (1.5,-1.25) .. (2,-1.25);

\draw[-] (2,0.75) ellipse (0.25 and 0.5);
\draw[-] (2,-0.75) ellipse (0.25 and 0.5);
\end{tikzpicture}
}\qquad\hspace{-0.2cm}
\subfloat[Copants\label{figs:copants}]
{
\begin{tikzpicture}[scale=\globalscale]

\node at (-0.25,0) {};
\node at (2.25,0) {};
\draw[-] (0,0) ellipse (0.25 and 0.5);

\draw[-] (0,0.5) .. controls (1,0.5) and (1.5,1.25) .. (2,1.25);
\draw[-] (2,0.25) .. controls (1.5,0) .. (2,-0.25);
\draw[-] (0,-0.5) .. controls (1,-0.5) and (1.5,-1.25) .. (2,-1.25);

\draw[-] (2,0.75) ellipse (0.25 and 0.5);
\draw[-] (2,-0.75) ellipse (0.25 and 0.5);
\end{tikzpicture}
}\qquad\hspace{-0.2cm}
\subfloat[Swap\label{figs:swap}]
{
\begin{tikzpicture}[scale=\globalscale]
\draw[-] (0,-0.75) ellipse (0.25 and 0.5);
\draw[-] (0,0.75) ellipse (0.25 and 0.5);

\draw[-,dashed,very thin] (0,0.25) .. controls (0.5,0.25) and (1.5,-1.25) .. (2,-1.25);
\draw[-,dashed,very thin] (0,1.25) .. controls (0.5,1.25) and (1.5,-0.25) .. (2,-0.25);

\draw[-] (0,1.25) .. controls (0.25,1.225) and (0.5,1) .. (1,0.51);
\draw[-] (0,0.25) .. controls (0.25,0.245) and (0.5,-0.05) .. (0.47,0);

\draw[-] (1.53,0) .. controls (1.5,0.05) and (1.75,-0.245) .. (2,-0.25);
\draw[-] (2,-1.25) .. controls (1.75,-1.225) and (1.5,-1) .. (1,-0.51);

\draw[-] (0,-0.25) .. controls (0.5,-0.25) and (1.5,1.25) .. (2,1.25);
\draw[-] (0,-1.25) .. controls (0.5,-1.25) and (1.5,0.25) .. (2,0.25);

\draw[-] (2,-0.75) ellipse (0.25 and 0.5);
\draw[-] (2,0.75) ellipse (0.25 and 0.5);
\end{tikzpicture}
}
\caption{The generators of all bordisms in $d = 2$.}
\label{figs:decomposition}
\end{figure}

This algebraic interpretation proceeds as follows. The key simplification in $d = 2$ is that the category 2\textbf{Cob} admits a description in terms of generators and relations, since surfaces can be completely classified. Namely, any 2-bordism can be decomposed in terms of a finite set of simple bordisms that generate all other bordisms by composition. The generators of 2\textbf{Cob} are shown in Figure \ref{figs:decomposition}---we refer to them as the cup, cap, cylinder, pants, copants, and swap. A 2D TQFT is completely determined by specifying the linear maps that it associates to the bordisms in this generating set.

\begin{figure}
\centering
\input{tkz/frobAlg.tex}
\caption{The equivalence relations of $2\textbf{Cob}$. Under the TQFT functor, these become the defining relations of a commutative Frobenius algebra on the Hilbert space of the TQFT on a circle.}
\label{figs:frobAlg}
\end{figure}

Since bordisms are regarded as equivalent up to homeomorphism, these generators obey various relations. The relations in 2\textbf{Cob} are summarized in Figure \ref{figs:frobAlg}. In anticipation of their algebraic nature, we refer to these relations as the (co)unit relations (Figures \ref{figs:frobAlg1} and \ref{figs:frobAlg2}, right), (co)associativity (Figures \ref{figs:frobAlg1} and \ref{figs:frobAlg2}, left), (co)commutativity (Figure \ref{commutativity}), and the \emph{Frobenius identity} (Figure \ref{figs:frobAlg3}).\footnote{These relations are not minimal---the Frobenius and (co)unit relations imply the (co)associativity relations \cite{Kock:2004cob}.} For brevity, we omit relations that involve sewing cylinders onto the other generators, as well as relations between disconnected bordisms involving swaps \cite{Kock:2004cob}.

\begin{table}
\centering
\begin{equation*}
\arraycolsep=10pt
\begin{array}{c|c|c}
\text{Bordism} & Z(\text{Bordism}) & \text{Algebraic Operation} \\
\hline \hline
\text{cylinder} & \text{id}_{\mathcal{H}_{S^1}}: \mathcal{H}_{S^1} \to \mathcal{H}_{S^1} & \text{identity} \\
\text{cup} & \eta: \mathbb{C} \to \mathcal{H}_{S^1} & \text{unit} \\
\text{cap} & \varepsilon: \mathcal{H}_{S^1} \to \mathbb{C} & \text{counit} \\
\text{pants} & \mu: \mathcal{H}_{S^1} \otimes \mathcal{H}_{S^1} \to \mathcal{H}_{S^1} & \text{multiplication} \\
\text{copants} & \delta: \mathcal{H}_{S^1} \to \mathcal{H}_{S^1} \otimes \mathcal{H}_{S^1} & \text{comultiplication} \\
\text{swap} & \tau: \mathcal{H}_{S^1} \otimes \mathcal{H}_{S^1} \to \mathcal{H}_{S^1} \otimes \mathcal{H}_{S^1} & \text{swap}
\end{array}
\end{equation*}
\caption{Elementary bordisms in 2\textbf{Cob}, their images under the TQFT functor $Z$, and their Frobenius algebra interpretation. The swap operator is the isomorphism of 2-party Hilbert spaces that takes $\ket{i} \otimes \ket{j} \mapsto \ket{j} \otimes \ket{i}$. A unitary 2D TQFT is one for which $\varepsilon = \eta^\dagger$ and $\delta = \mu^\dagger$.}
\label{elementarybordisms}
\end{table}

We list the images of the generators in 2\textbf{Cob} under the TQFT functor in Table \ref{elementarybordisms}. The complex vector space $\mathcal{H}_{S^1}$ is said to be a \emph{Frobenius algebra} over $\mathbb{C}$ if there exist linear maps as listed in Table \ref{elementarybordisms} that satisfy the following relations:
\begin{itemize}
\item[(1)] \textit{associative unital multiplication}:
\begin{align}
\mu \circ (\text{id}_{\mathcal{H}_{S^1}} \otimes \mu) &= \mu \circ (\mu \otimes \text{id}_{\mathcal{H}_{S^1}}),\\
\mu \circ (\text{id}_{\mathcal{H}_{S^1}} \otimes \eta) &= \text{id}_{\mathcal{H}_{S^1}} = \mu \circ (\eta \otimes \text{id}_{\mathcal{H}_{S^1}}),
\end{align}

\item[(2)] \textit{coassociative counital comultiplication}:
\begin{align}
(\text{id}_{\mathcal{H}_{S^1}} \otimes \delta) \circ \delta &= (\delta \otimes \text{id}_{\mathcal{H}_{S^1}}) \circ \delta,\\
(\text{id}_{\mathcal{H}_{S^1}} \otimes \varepsilon) \circ \delta &= \text{id}_{\mathcal{H}_{S^1}} = (\varepsilon \otimes \text{id}_{\mathcal{H}_{S^1}}) \circ \delta,
\end{align}

\item[(3)] \textit{Frobenius identity}:
\begin{equation}
(\mu \otimes \text{id}_{\mathcal{H}_{S^1}}) \circ (\text{id}_{\mathcal{H}_{S^1}} \otimes \delta) = \delta \circ \mu = (\text{id}_{\mathcal{H}_{S^1}} \otimes \mu) \circ (\delta \otimes \text{id}_{\mathcal{H}_{S^1}}).
\end{equation}
\end{itemize}
$\mathcal{H}_{S^1}$ is a \emph{commutative} Frobenius algebra if, in addition to the above, we have
\begin{equation}
\mu \circ \tau = \mu.\label{comm}
\end{equation}
A commutative Frobenius algebra is automatically cocommutative ($\tau \circ \delta = \delta$) \cite{Kock:2004cob}. In light of the relations in $2\textbf{Cob}$ (Figure \ref{figs:frobAlg}), we see that a 2D TQFT imbues its $S^1$ Hilbert space $\mathcal{H}_{S^1}$ with the structure of a commutative Frobenius algebra.

It is useful to have a more economical definition of a Frobenius algebra. To that end, let $A$ be a finite-dimensional algebra over $\mathbb{C}$. Recall that any linear functional $\varepsilon : A\to \mathbb{C}$ canonically determines an associative bilinear form $\sigma : A\times A\to \mathbb{C}$ (via $\sigma(x, y) = \varepsilon(xy)$) and vice versa (via $\varepsilon(x) = \sigma(1_A, x) = \sigma(x, 1_A)$), where associativity means $\sigma(xa, y) = \sigma(x, ay)$. Then:
\begin{quote}
\textit{A \emph{Frobenius algebra} is a finite-dimensional $\mathbb{C}$-algebra $A$ equipped with a linear functional $\varepsilon$ (the \emph{counit}) such that the corresponding associative bilinear form $\sigma$ (the \emph{Frobenius form}) is nondegenerate.}\footnote{The terminology for $\varepsilon$ comes from an alternative characterization of a Frobenius algebra as an algebra that is simultaneously a coalgebra, with a compatibility condition between multiplication and comultiplication given by the Frobenius identity.}
\end{quote}
In terms of the associated linear map $\sigma : A\otimes A\to \mathbb{C}$, the Frobenius form is defined as
\begin{equation}
\sigma = \varepsilon \circ \mu,
\end{equation}
where $\mu : A\otimes A\to A$ is the multiplication operation on $A$. The nondegeneracy of the Frobenius form guarantees the existence of a dual coform $\gamma: \mathbb{C} \to A\otimes A$, which allows one to construct the comultiplication operation as $\delta = (\text{id}_A \otimes \mu) \circ (\gamma(1) \otimes \text{id}_A)$.

In 2D TQFT, $\mathcal{H}_{S^1}$ has the structure of a commutative, associative algebra with multiplication given by the pair-of-pants operation in the path integral. Specifying a counit (cap) further yields a nondegenerate bilinear form $\sigma : \mathcal{H}_{S^1}\otimes \mathcal{H}_{S^1}\to \mathbb{C}$, or ``U-tube'' (pants composed with cap operation). This gives $\mathcal{H}_{S^1}$ the structure of a Frobenius algebra, which is moreover commutative. When we write $\mathcal{H}_{S^1} = A$ where $A$ is an algebra over $\mathbb{C}$, we mean that $\mathcal{H}_{S^1}\cong A$ as a $\mathbb{C}$-vector space and that the corresponding 2D TQFT is defined by the Frobenius algebra obtained from $A$ by specifying a counit (or equivalently, a Frobenius form).

%%%%%%%%%%%%%%%%%%%%%%%%%%%%%%%%%%%%%%%%%%%%%%%
\subsubsection{Classification} \label{subsubsec:classification}
%%%%%%%%%%%%%%%%%%%%%%%%%%%%%%%%%%%%%%%%%%%%%%

A basic observation is that Frobenius structures are compatible with direct sum. Given two Frobenius algebras $(A', \varepsilon')$ and $(A'', \varepsilon'')$, their direct sum $(A, \varepsilon)$ is defined by $A = A'\oplus A''$ and $\varepsilon : a'\oplus a''\mapsto \varepsilon'(a') + \varepsilon''(a'')$, where nondegeneracy of $\varepsilon'$ and $\varepsilon''$ implies nondegeneracy of $\varepsilon$. For 2D TQFTs, the direct sum operation of \cite{Durhuus:1993cq} corresponds simply to the direct sum of Frobenius algebras \cite{Sawin:1995rh}.

Any 2D TQFT can be written as a direct sum of two types of theories: \textit{nilpotent} and \textit{semisimple}. We define them as follows \cite{Sawin:1995rh}.

For each nonzero $z\in \mathbb{C}$, let $S_z$ denote the Frobenius algebra $\mathbb{C}$ with $\varepsilon$ being multiplication by $z^{-1}$. Each $S_z$ furnishes a distinct 2D TQFT with 1D Hilbert space on $S^1$. Any indecomposable commutative Frobenius algebra with no nilpotent elements is isomorphic to $S_z$ for some $z$, and we call such a 2D TQFT \emph{simple}. On the other hand, we call a 2D TQFT \emph{nilpotent} if, as a commutative Frobenius algebra, it is indecomposable and contains at least one nilpotent element (see \cite{Sawin:1995rh} for details of their characterization).\footnote{Nilpotent 2D TQFTs satisfy the property that any bordism of genus two or higher corresponds to the zero operator on the appropriate space \cite{Sawin:1995rh}.} Then we have the following basic classification result:
\begin{center}
\emph{Every 2D TQFT is a direct sum of simple and nilpotent theories.}
\end{center}
In particular, every indecomposable commutative $H^\ast$-algebra takes the form $S_\lambda$ for some positive real $\lambda$, and every unitary 2D TQFT can be written as a direct sum of such theories $S_\lambda$ \cite{Durhuus:1993cq}.

We refer to a direct sum of simple TQFTs as \emph{semisimple}; such theories will be our primary focus. Semisimple 2D TQFTs are precisely those in which the \emph{handle operator}
\begin{equation}
\mu \circ \delta: \mathcal{H}_{S^1} \to \mathcal{H}_{S^1}
\end{equation}
(the linear operator corresponding to composition of pants with copants) is diagonalizable. In a unitary 2D TQFT, the handle operator is diagonalizable by virtue of being Hermitian \cite{Durhuus:1993cq}.

All 2D TQFTs with 2D Hilbert space on a circle may be classified up to isomorphism \cite{Fenyes:2015frb}, as we review in Appendix \ref{app:frobenius}.  There, we give explicit representations of the linear maps in Table \ref{elementarybordisms} in all possible cases.

%%%%%%%%%%%%%%%%%
\section{Induced Complexity in TQFT}\label{sec:inducedComplexity}
%%%%%%%%%%%%%%%%%

We now define complexity in the context of functorial 2D TQFT. Our main claim is that there exists a simple and natural notion of complexity on 2\textbf{Cob} arising from the pants decomposition, and that this definition of complexity induces one on \textbf{Hilb} via the TQFT functor. In physical terms, the complexity of the background topology for the TQFT path integral should determine the complexity of the corresponding states in the TQFT Hilbert space and that of the linear maps that prepare them from some reference state.

However, this na\"ive definition of complexity is clearly deficient. The TQFT functor from 2\textbf{Cob} to \textbf{Hilb} need not be surjective: in general, not all linear maps between Hilbert spaces can be realized as the image of some bordism. Moreover, the set of linear maps that can be realized in this way need not be universal. Put another way, the TQFT path integral generally cannot prepare a set of states that is anywhere near dense in the boundary Hilbert space. Therefore, despite that 2\textbf{Cob} contains a universal set of elementary gates (morphisms), the image of this universal gate set under the TQFT functor is not universal for \textbf{Hilb}. This would seem to be a major obstacle to defining induced operator and state complexity on \textbf{Hilb}.

We will partially surmount these difficulties by showing that in semisimple 2D TQFTs, the TQFT functor induces a ``weak'' form of universality on \textbf{Hilb}, thereby realizing a one-to-one correspondence between bordism complexity and state complexity for a specific class of states.

%%%%%%%%%%%%%%%%%
\subsection{Categorical Definition}\label{subsec:categoricaldef}
%%%%%%%%%%%%%%%%%

We first define complexity in 2\textbf{Cob}. This is a relatively simple task because the morphisms of 2\textbf{Cob} are finitely generated by the bordisms in Figure \ref{figs:decomposition}, which comprise an exactly universal gate set. The complexity of a given 2-bordism is then the minimum number of elementary gates required to build it.

In fact, for a generic bordism, we may ignore the cup, cap, cylinder, and swap generators altogether and simply quantify the circuit complexity using the \emph{pants decomposition}. Namely, the number of pairs of pants into which we may divide a connected bordism of genus $g$ with $m$ boundary components is
\begin{equation}
\text{\# pants} = 2g - 2 + m = -\chi(g,m),
\label{numPants}
\end{equation}
where $\chi(g,m)$ is the Euler characteristic.\footnote{This formula holds for a connected surface with negative Euler characteristic, i.e., one for which a pants decomposition exists. On such a surface, the maximum number of simple closed curves that are disjoint, homotopically distinct, and homotopic to neither a point nor a boundary component is $3g - 3 + m$, and a collection of such curves cuts the surface into $2g - 2 + m$ pairs of pants \cite{Ratcliffe1994}.} (Our usage of the word ``pants'' here ignores orientation, and hence includes both the pants and copants generators in Figures \ref{figs:pants} and \ref{figs:copants}, respectively.)

To phrase this definition of complexity in a way that will be more useful when passing to \textbf{Hilb}, we recall some facts about surfaces. Two connected, compact, oriented surfaces with oriented boundary are homeomorphic if and only if they have the same genus and the same number of in- and out-boundaries. Moreover, the only invertible 2-bordisms are the \emph{permutation bordisms} built solely from cylinders and swaps, since the genus is additive under composition of bordisms. Every connected 2-bordism can be expressed as a composition and disjoint union of the five non-swap generators. Every 2-bordism (possibly disconnected) factors as a disjoint union of connected 2-bordisms, preceded and followed by permutation bordisms. We focus on connected bordisms for simplicity.

We now introduce the \textit{normal form} \cite{Kock:2004cob} of a connected bordism, as sketched in Figure \ref{fig:cobMap}. This is a canonical decomposition into generators in which the surface ``fans in'' to a single $S^1$ from the in-boundaries, decomposes into a sequence of handles, and then ``fans out'' to the out-boundaries. If there are no in- or out-boundaries, then the in- or out-part is a cup or a cap. Algebraically, this decomposition entails writing the linear map corresponding to an arbitrary connected bordism from $(S^1)^p$ to $(S^1)^q$ as
\begin{equation}
\mathcal{O}^{p,q}_{g} = \Delta_q \circ (\mu \circ \delta)^g \circ M_p,\label{genOp}
\end{equation}
where $M_p: \mathcal{H}_{S^1}^{\otimes p} \to \mathcal{H}_{S^1}$ denotes $p$-fold multiplication, $\Delta_q: \mathcal{H}_{S^1} \to \mathcal{H}_{S^1}^{\otimes q}$ denotes $q$-fold comultiplication, and $\mu \circ \delta: \mathcal{H}_{S^1} \to \mathcal{H}_{S^1}$ is the handle operator. We define $M_0 = \eta$ and $\Delta_0 = \varepsilon$, as well as $M_1 = \Delta_1 = \text{id}_{\mathcal{H}_{S^1}}$.  For $p,q \geq 1$, we define $M_p$ and $\Delta_q$ recursively by
\begin{align}
M_{p+1} &= M_{p} \circ \underbrace{(\mu \otimes \text{id}_{\mathcal{H}_{S^1}} \otimes \cdots \otimes \text{id}_{\mathcal{H}_{S^1}})}_{p\ \text{factors}},\\
\Delta_{q+1} &= {\underbrace{(\text{id}_{\mathcal{H}_{S^1}} \otimes \cdots \otimes \text{id}_{\mathcal{H}_{S^1}} \otimes \delta)}_{q\ \text{factors}}} \circ \Delta_{q}.
\end{align}
Of course, we could simply define the complexity of the linear operator $\mathcal{O}^{p,q}_{g}$ in \textbf{Hilb} to be the complexity of its preimage in 2\textbf{Cob} according to \eqref{numPants}, regardless of whether we present the corresponding bordism in normal form:\footnote{Again, this formula holds up to edge cases (small $g, p, q$) because it assumes that the bordism can be decomposed solely into (co)pants.  It fails when $g = 0$ and $p + q\leq 2$, or $g = 1$ and $(p, q) = (0, 0)$.}
\begin{equation}
\mathcal{C}(\mathcal{O}_g^{p,q}) = 2g - 2 + p + q.
\label{normSize}
\end{equation}
But the essential point of normal form is that for fixed $p, q$, the complexity of a connected bordism from $(S^1)^p$ to $(S^1)^q$ is determined solely by the genus $g$ (the operations of fanning in or out incur only a constant overhead). In this sense, the only computationally useful gate in 2\textbf{Cob} is the handle bordism from $S^1$ to $S^1$ (which is a \emph{composite} of two of the generators in Figure \ref{figs:decomposition}), and likewise, the handle operator $\mu \circ \delta$ should be regarded as the only elementary gate in \textbf{Hilb} for the purposes of complexity in 2D TQFT. The complexity of a morphism in \textbf{Hilb} that lies in the image of the TQFT functor is then determined by the number of occurrences of this operator.

To illustrate these considerations, we next compute the 1-party states attainable from the TQFT path integral in two example theories, as well as their corresponding complexities. For state complexity in \textbf{Hilb}, we choose as our reference state the image of the cup $\eta(1)$, which is the simplest state that can be constructed directly from the TQFT path integral.

%%%%%%%%%%%%%%%%%
\subsubsection{Nilpotent Example: \texorpdfstring{$\mathbb{C}[x]/(x^n)$}{C[x]/(x**n)}}\label{subsubsec:nilpotentEx}
%%%%%%%%%%%%%%%%%

The prototypical example of a nilpotent 2D TQFT has $\mathcal{H}_{S^1}\cong \mathbb{C}[x]/(x^n)$ ($n > 1$), with counit $x^{n-1}\mapsto 1$ and $x^i\mapsto 0$ for $i < n - 1$ \cite{Kock:2004cob, Peelaers:2020ncm}.

We denote the basis states $x^i$ in Dirac notation by
\begin{equation}
\ket{i}, \quad i = 0, \ldots, n - 1.
\end{equation}
This vector space has the structure of a commutative Frobenius algebra with unit, counit, multiplication, and comultiplication defined as
\begin{align}
&\eta(1) = \ket{0},&
&\mu(\ket{i}\otimes \ket{j}) = \delta_{i + j < n}\ket{i+j}, \label{multxn} \\
&\varepsilon(\ket{i}) = \delta_{i, n - 1},&
&\delta(\ket{i}) = \textstyle \sum_{j=i}^{n-1} \ket{n-1+i-j} \otimes \ket{j}, \label{comultxn}
\end{align}
where we have introduced a generalized Kronecker delta function:
\begin{equation}
\delta_P\equiv \begin{cases} 1 & \text{if $P$ is true}, \\ 0 & \text{if $P$ is false}.\end{cases}
\end{equation}
The generalized $p$-fold multiplication map $M_p: \mathcal{H}_{S^1}^{\otimes p} \to \mathcal{H}_{S^1}$ ($p \geq 2$) is
\begin{equation}
M_p(\ket{i_1} \otimes \cdots \otimes \ket{i_p}) = \delta_{\sum_{k=1}^{p} i_k < n}\ket{\textstyle \sum_{k=1}^{p} i_k}, \label{pfoldxn}
\end{equation}
and the generalized $q$-fold comultiplication map $\Delta_q: \mathcal{H}_{S^1} \to \mathcal{H}_{S^1}^{\otimes q}$ ($q \geq 2$) is
\begin{equation}
\begin{split}
\Delta_{q}(\ket{i}) &= \sum_{j_1 = 0}^{n-1}\sum_{j_2 = j_1}^{n-1}\cdots \sum_{j_q = j_{q-1}}^{n-1} \delta_{i, j_1}\ket{n - 1 + j_1 - j_2}\otimes \cdots\\[-3ex]
&\quad\qquad\qquad\qquad\qquad\qquad\otimes \ket{n - 1 + j_{q-1} - j_q}\otimes \ket{j_q}.
\end{split}
\label{qfoldxn}
\end{equation}
To a closed, connected surface of genus $g$, this TQFT associates the invariant $\dim\mathcal{H}_{S^1} = n$ if $g = 1$ and 0 otherwise, as can be seen from the nilpotence of the corresponding handle operator.

To examine state complexity, we must know which states can be constructed from the TQFT path integral. We first determine the states that can be prepared in $\mathcal{H}_{S^1}$ via the path integral on connected surfaces with $S^1$ boundary. The unit itself allows us to construct $\ket{0}$. Using \eqref{multxn} and \eqref{comultxn}, we find that
\begin{equation}
\mathcal{O}_g^{1,1}(\ket{0}) = (\mu \circ \delta)^g (\ket{0}) = \begin{cases} \ket{0} & \text{if $g = 0$}, \\ n\ket{n-1} & \text{if $g = 1$}, \\ 0 & \text{if $g\geq 2$}. \end{cases}
\end{equation}
Similarly, the states that can be prepared in $\mathcal{H}_{S^1}^{\otimes q}$ using the TQFT path integral on connected surfaces with $(S^1)^q$ boundary are the following:
\begin{equation}
\mathcal{O}_g^{1, q\geq 2}(\ket{0}) = \Delta_q\circ (\mu \circ \delta)^g (\ket{0}) = \begin{cases} \Delta_q(\ket{0}) & \text{if $g = 0$}, \\ n\ket{n-1}^{\otimes q} & \text{if $g = 1$}, \\ 0 & \text{if $g\geq 2$}. \end{cases}
\end{equation}
Acting on $\ket{0}$ with an operator $\mathcal{O}_g^{1, q}$ for which $g = 0$ and $q \geq 2$ produces entangled $q$-party states. Restricting our attention to the accessible 1-party states in $\mathcal{H}_{S^1}$, all of which may be prepared by connected bordisms, we conclude that only the basis states $\ket{0}$ and $\ket{n - 1}$ have finite complexity. That is, the state complexity of a 1-party state $\ket{\psi}$ with respect to the reference $\ket{0}$ (as measured by the pants decomposition of the bordism that prepares it) is
\begin{equation}
\mathcal{C}_{\ket{0}}(\ket{\psi}) = \begin{cases}
0 & \text{if $\ket{\psi} \sim \ket{0}$}, \\
2 & \text{if $\ket{\psi} \sim \ket{n-1}$}, \\
\infty & \text{otherwise}.
\end{cases}
\end{equation}
Above, $\sim$ denotes equivalence in the projective Hilbert space $\mathbb{P}(\mathcal{H}_{S^1}) \cong \mathbb{CP}^{n-1}$. These complexities are a manifestation of the generic lack of universality in 2D TQFT.

%%%%%%%%%%%%%%%%%
\subsubsection{Semisimple Example: \texorpdfstring{$\mathbb{C}[x]/(x^n - 1)$}{C[x]/(x**n - 1)}}\label{subsubsec:semisimpleEx}
%%%%%%%%%%%%%%%%%

The prototypical example of a semisimple 2D TQFT has $\mathcal{H}_{S^1}\cong \mathbb{C}[x]/(x^n - 1)$ (the group algebra of $\mathbb{Z}_n$), with counit $1\mapsto 1$ and $x^i\mapsto 0$ for $i > 0$ \cite{Kock:2004cob, Peelaers:2020ncm}.

Again, we denote the basis states $x^i$ by $\ket{i}$ with $i = 0, \ldots, n - 1$. The unit, counit, multiplication, and comultiplication are given by
\begin{align}
&\eta(1) = 0, &
&\mu(\ket{i} \otimes \ket{j}) = \ket{(i + j)\text{ mod } n}, \\
&\varepsilon(\ket{i}) = \delta_{i, 0}, &
&\delta(\ket{i}) = \textstyle \sum_{j=0}^{n-1} \ket{(i - j)\text{ mod } n} \otimes \ket{j},
\end{align}
while $p$-fold multiplication and $q$-fold comultiplication (for $p,q \geq 2$) take the form
\begin{align}
&M_p(\ket{i_1} \otimes \cdots \otimes \ket{i_p}) = \ket{\textstyle (\sum_{k=1}^{p} i_k)\text{ mod } n}, \\
&\Delta_q(\ket{i}) = \sum_{j_1, \ldots, j_q = 0}^{n-1} \delta_{i, j_1}\ket{(j_1 - j_2)\text{ mod } n}\otimes \cdots\otimes \ket{(j_{q-1} - j_q)\text{ mod } n}\otimes \ket{j_q}.
\end{align}
In this case, the handle operator is simply multiplication by $n$, so this TQFT associates to a closed, connected surface of genus $g$ the invariant $n^g$.

The states that can be prepared from the reference $\ket{0}$ via connected bordisms take the form
\begin{equation}
(\mu \circ \delta)^g(\ket{0}) = n^g \ket{0}, \qquad \mathcal{O}_{g}^{1,q\geq 2}(\ket{0}) = n^g\Delta_q(\ket{0}),
\end{equation}
where the $q$-party states are entangled for any $g$. The prefactor of $n^g$ is irrelevant in the projective Hilbert space, so the only accessible 1-party state in this example is $\ket{0}$, with complexity 0.

%%%%%%%%%%%%%%%%%%%%%%%%%%%%%%%
\subsection{Universality}\label{subsec:universality}
%%%%%%%%%%%%%%%%%%%%%%%%%%%%%%%

The 2D TQFTs considered above lack even approximate universality on 1-party states, and more generally on \textbf{Hilb}, despite the presence of an exactly universal gate set in $2\textbf{Cob}$. These examples highlight that the universality of the generators in $2\textbf{Cob}$ fails to be preserved under the TQFT functor.

Without universality, complexity as a descriptor lacks any nuance. We would say that non-constructible circuits and states have infinite complexity. To develop a good notion of complexity for TQFT, we ask: which 2D TQFTs, if any, allow for a positive-dimensional manifold of (approximately) constructible states? In this scenario, there would exist states with arbitrarily large but finite complexities.

This question has a simple answer: even though the TQFT path integral is generically non-universal, for certain TQFTs in which the handle operator has multiple (nonzero) eigenvalues of largest norm, the path integral is capable of preparing to arbitrary accuracy all states in a \emph{toroidal} submanifold of Hilbert space. For states in this submanifold, the complexity of state preparation is meaningful and correlates precisely with the topological complexity of the path integral (either a minimum number of handles or a minimum number of pairs of pants).

To illustrate, consider a 2D TQFT with $\dim\mathcal{H}_{S^1} = n$, and suppose that the handle operator $H\equiv \mu\circ \delta$ may be diagonalized into the form
\begin{equation}
H = \operatorname{diag}(e^{i\phi_0}, \ldots, e^{i\phi_{n-1}}), \quad \phi_i \in \mathbb{R},
\end{equation}
up to an overall factor. In this case, all of the eigenvalues have equal magnitude. Let $(c_0, \ldots, c_{n-1})$ denote the coefficients of a generic reference state $\ket{\psi_0}$ in the eigenbasis of $H$ ($c_i\in \mathbb{C}$). Using obvious shorthand, applying the handle operator $g$ times prepares the state
\begin{equation}
H^g \ket{\psi_0} = (c_0 e^{ig\phi_0}, \ldots, c_{n-1}e^{ig\phi_{n-1}}).
\end{equation}
For generic phases $\phi_i$ (meaning incommensurate, irrational multiples of $2\pi$), there exists some $g$ for which $H^g \ket{\psi_0}$ approximates any target state of the form
\begin{equation}
(c_0 e^{i\theta_0}, \ldots, c_{n-1}e^{i\theta_{n-1}}), \quad \theta_i\in \mathbb{R}
\end{equation}
to within any tolerance $\epsilon$. Universality for this subset of states is guaranteed by the ergodicity of irrational rotations, and the complexity reflects the \emph{hitting time} for an interval corresponding to an $\epsilon$-ball around the target state.

Since the overall phase is irrelevant, the set of states that can be prepared exactly is dense in a real $(n - 1)$-dimensional torus $\mathbb{T}^{n-1}$ inside the projective 1-party Hilbert space $\mathbb{P}(\mathcal{H}_{S^1}) \cong \mathbb{CP}^{n-1}$, parametrized by $n - 1$ independent phase angles. This torus has half the dimension of the projective Hilbert space.

There are various special cases that one could consider. If one of the phases $\phi_i$ is a rational multiple of $2\pi$ for some reduced fraction $p/q$, then universality persists as long as we choose the tolerance sufficiently large relative to $1/q$. Moreover, one can start with a state where one of the amplitudes $c_i$ is zero, leading us to access a torus of dimension less than $n - 1$. Something similar happens if at least one pair of the phases is commensurate.

The above discussion concerns 1-party states. Restricting to reference states that can be prepared by the TQFT path integral, this discussion extends to $N$-party states by considering both connected and disconnected bordisms from $\varnothing$ to $(S^1)^N$. The $N$-party states that are accessible via connected bordisms are obtained by applying all necessary handle operators at the 1-party level, followed by $q$-fold comultiplication. ``Disconnected'' states in $N$ parties are obtained by tensoring connected states of fewer parties. We may restrict our attention to bordisms whose number of connected components does not exceed the number of parties, since connected components with no boundaries correspond to multiplication by an overall factor.

How much of the $N$-party Hilbert space can the TQFT path integral access?  In the generic case, the accessible 1-party states comprise a torus $\mathbb{T}^{n-1}$.  Up to permutation bordisms, the accessible $N$-party states are classified by the integer partitions of $N$, and the dimensionality of the submanifold that they occupy is determined by the number of parts in the partition.  Clearly, the ``maximally'' disconnected $N$-party states ($N$-fold tensor products of $1$-party states) maximize the number of free parameters.  Such states comprise an $N(n-1)$-dimensional torus $\mathbb{T}^{N(n-1)}$ inside the $N$-party projective Hilbert space $\mathbb{P}(\mathcal{H}_{S^1}^{\otimes N}) \cong \mathbb{CP}^{n^N - 1}$. The ratio
\begin{equation}
\frac{\dim\mathbb{T}^{N(n-1)}}{\dim\mathbb{CP}^{n^N-1}} = \frac{N(n-1)}{2(n^N - 1)}
\label{ratio}
\end{equation}
decreases \emph{exponentially} with $N$, starting at $1/2$ for $N = 1$. The situation is only worse for other disconnected states, obtained by considering connected components with more than one party: such states occupy lower-dimensional tori. Hence the relative size of the maximal constructible state space within the entire Hilbert space decreases with the number of parties $N$. Even for those TQFTs with some degree of universality, the dimension of the state space that the path integral can access is limited to no more than half of the Hilbert space dimension.

We now sharpen some of the preceding points, first in the context of TQFTs with $n = 2$ and then in the context of general semisimple TQFTs.

%%%%%%%%%%%%%%%%%%%%%%%%%%%%%%%
\subsubsection{``Two-Level'' TQFTs}\label{subsubsec:twolevel}
%%%%%%%%%%%%%%%%%%%%%%%%%%%%%%%

When $n = 2$, the previous discussion can be made very concrete using the classification of 2D Frobenius algebras presented in Appendix \ref{app:frobenius}. For 2D TQFTs with $n = 2$, we determine the toroidal submanifolds of constructible states and illustrate the $N$-party dimension counting discussed above. Such TQFTs fall into two classes, which we refer to as type I or type II depending on whether the Frobenius form satisfies $\sigma(\ket{00}) = 0$ or $\sigma(\ket{00}) \neq 0$, respectively, where $\eta(1)\equiv \ket{0}$ is the unit for $\mu$. We define the state $\ket{1}$ such that $\{\ket{0}, \ket{1}\}$ is an orthonormal basis for $\mathcal{H}_{S^1}$. TQFTs of type I are specified by a single complex number $c\in \mathbb{C}$, while TQFTs of type II are specified by two complex numbers $c\in \mathbb{C}$ and $d\in \mathbb{C}\setminus \{0\}$.

We first discuss the 1-party states that can be prepared by the TQFT path integral, which amounts to computing the action of powers of the handle operator $\mu \circ \delta$.

For TQFTs of type I, the handle operator is the following linear operator on $\mathcal{H}_{S^1}$ in the $\{\ket{0}, \ket{1}\}$ basis:
\begin{equation}
\mu \circ \delta = 2\begin{pmatrix}
0 && c\\
1 && 0
\end{pmatrix}.
\label{t1Handle}
\end{equation}
Hence the attainable 1-party states are
\begin{equation}
[(\mu\circ\delta)^g\circ \eta](1) = 2^g c^{\lfloor g/2\rfloor}\ket{g\text{ mod } 2}.
\label{type1Im}
\end{equation}
For $c\neq 0$, the boundary state alternates between $\ket{0}$ and $\ket{1}$ in the projective Hilbert space $\mathbb{CP}^1$ as we add handles.  Equivalently, assuming $c\neq 0$, we have in terms of the unnormalized eigenvectors $\ket{v_{\pm}} = \pm\sqrt{c}\ket{0} + \ket{1}$ with eigenvalues $\lambda_{\pm} = \pm 2\sqrt{c}$ that
\begin{equation}
(\mu \circ \delta)^g \ket{0} \sim \lambda_+^g \ket{v_+} - \lambda_-^g \ket{v_-} \sim \ket{v_+} - (-1)^g \ket{v_-}.
\end{equation}
Thus type I TQFTs can prepare only a finite set of 1-party states.

On the other hand, for TQFTs of type II, we have
\begin{equation}
\mu \circ \delta = \frac{1}{d}\begin{pmatrix}
2 && c\\
c && 2+c^2
\end{pmatrix}
\label{t2Handle}
\end{equation}
in the $\{\ket{0}, \ket{1}\}$ basis. Iterating this matrix, we obtain
\begin{equation}
[(\mu \circ \delta)^g \circ \eta](1) = \frac{\lambda_+^{g-1} + \lambda_-^{g-1}}{d}\ket{0} + \frac{\lambda_+^g - \lambda_-^g}{\sqrt{4 + c^2}}\ket{1},
\end{equation}
where the eigenvalues and corresponding eigenvectors of the handle operator are
\begin{equation}
\lambda_{\pm} = \frac{\sqrt{4 + c^2}}{2d}\left(\sqrt{4 + c^2} \pm c\right), \qquad \ket{v_{\pm}} = \frac{-c\pm \sqrt{4 + c^2}}{2}\ket{0} + \ket{1}.
\end{equation}
Assuming $c\neq \pm 2i$, we have in $\mathbb{CP}^1$ that
\begin{equation}
(\mu \circ \delta)^g \ket{0} \sim \lambda_+^g \ket{v_+} - \lambda_-^g \ket{v_-} \sim \left(\sqrt{4 + c^2} + c\right)^g\ket{v_+} - \left(\sqrt{4 + c^2} - c\right)^g\ket{v_-}.
\end{equation}
If the eigenvalues $\lambda_\pm$ have different magnitudes, then one of the eigenvectors $\ket{v_+}$ or $\ket{v_-}$ becomes an accumulation point for the action of the handle operator.  We are interested in the case that $|\lambda_+| = |\lambda_-|$.  This occurs precisely when
\begin{equation}
c^2 = 2(\cos\phi - 1), \quad \phi\in \mathbb{R},
\label{densec}
\end{equation}
in which case
\begin{equation}
(\mu \circ \delta)^g \ket{0} \sim \ket{v_+} - e^{ig\phi} \ket{v_-}.
\end{equation}
As long as $\phi$ is an irrational multiple of $2\pi$, the handle operator acting on $\ket{0}$ can be used to approximate any possible phase on $-\ket{v_-}$ to arbitrary precision, so the constructible 1-party states densely fill a circle. The corresponding $c$ is purely imaginary, so the TQFT is non-unitary (in the functorial sense discussed in Section \ref{subsec:general}).

Note that the handle operator fails to be diagonalizable, and that the TQFT is therefore nilpotent, precisely when $c = 0$ in the type I case and $c = \pm 2i$ in the type II case. The TQFT is semisimple otherwise, and it assigns to a closed manifold the invariant
\begin{equation}
[\varepsilon\circ (\mu\circ \delta)^g\circ \eta](1) = \lambda_+^{g-1} + \lambda_-^{g-1},
\end{equation}
where $\lambda_\pm$ are the nonzero eigenvalues of the handle operator.

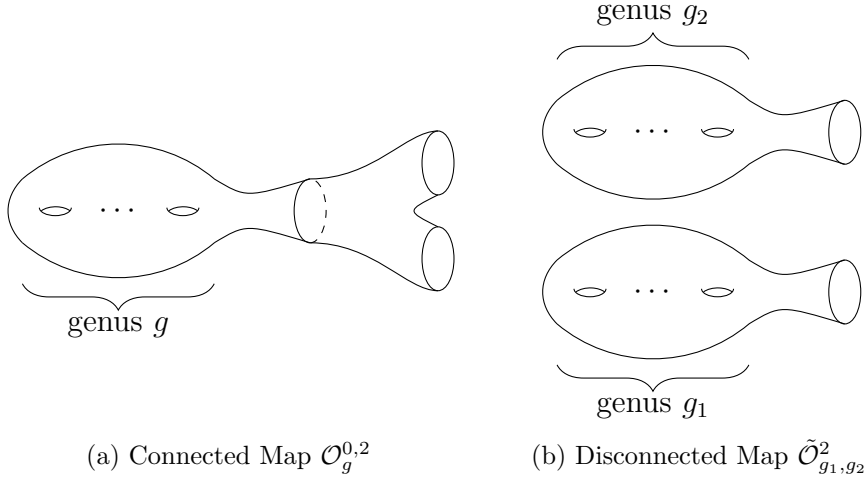
\begin{figure}
\centering
\subfloat[Connected Map $\mathcal{O}_{g}^{0,2}$\label{figs:connected}]
{
\begin{tikzpicture}[scale=0.85]
% Copants

\draw[-] (0+3.5,0.5) .. controls (1+3.5,0.5) and (1.5+3.5,1.25) .. (2+3.5,1.25);
\draw[-] (2+3.5,0.25) .. controls (1.5+3.5,0) .. (2+3.5,-0.25);
\draw[-] (0+3.5,-0.5) .. controls (1+3.5,-0.5) and (1.5+3.5,-1.25) .. (2+3.5,-1.25);

\draw[-] (2+3.5,0.75) ellipse (0.25 and 0.5);
\draw[-] (2+3.5,-0.75) ellipse (0.25 and 0.5);
%%%%%%%%%%%%%%%%%%%%%%%%%%%%%%%%%%%%%%%%%%%%%%%%%%%%%%%%%%%%%%%%%%%%

\draw[-,dashed] (2.5+1,-0.5) arc (270:450:0.25 and 0.5);
\draw[-] (2.5+1,-0.5) arc (270:90:0.25 and 0.5);

\draw[-] (2.5+1,0.5) .. controls (1.5+1,0.2)  .. (1+1,0.5);
\draw[-] (2.5+1,-0.5) .. controls (1.5+1,-0.2)  .. (1+1,-0.5);

\draw[-] (-1,0.5) .. controls (-1.3,0.2) and (-1.3,-0.2)  .. (-1,-0.5);

\draw[-] (-1,0.5) arc (130:50:1.56+0.775);
\draw[-] (-1,-0.5) arc (-130:-50:1.56+0.775);

\draw[-] (-0.735,0.05) arc (180:360:0.25 and 0.125);
\draw[-] (0.03-0.735,0.05-0.05) arc (145:35:0.275 and 0.15);

\draw[-] (0.525+0.735,0.05) arc (180:360:0.25 and 0.125);
\draw[-] (0.525+0.03+0.735,0.05-0.05) arc (145:35:0.275 and 0.15);

\node at (0.125-0.1+0.5,0) {$\cdots$};

\draw [decorate,decoration={brace,amplitude=10pt},yshift=-4pt,xshift=0pt]
(2,-1) -- (-1,-1);

\node at (0.5,-1.825) {genus $g$};

\node at (1,-3.25) {};
\end{tikzpicture}
}\quad\quad
\subfloat[Disconnected Map $\tilde{\mathcal{O}}_{g_1,g_2}^2$\label{figs:disconnected}]
{
\begin{tikzpicture}[scale=0.85]

\draw[-] (2.5+1,-0.5+2.5) arc (270:450:0.25 and 0.5);
\draw[-] (2.5+1,-0.5+2.5) arc (270:90:0.25 and 0.5);

\draw[-] (2.5+1,0.5+2.5) .. controls (1.5+1,0.2+2.5)  .. (1+1,0.5+2.5);
\draw[-] (2.5+1,-0.5+2.5) .. controls (1.5+1,-0.2+2.5)  .. (1+1,-0.5+2.5);

\draw[-] (-1,0.5+2.5) .. controls (-1.3,0.2+2.5) and (-1.3,-0.2+2.5)  .. (-1,-0.5+2.5);

\draw[-] (-1,0.5+2.5) arc (130:50:1.56+0.775);
\draw[-] (-1,-0.5+2.5) arc (-130:-50:1.56+0.775);

\draw[-] (-0.735,0.05+2.5) arc (180:360:0.25 and 0.125);
\draw[-] (0.03-0.735,0.05-0.05+2.5) arc (145:35:0.275 and 0.15);

\draw[-] (0.525+0.735,0.05+2.5) arc (180:360:0.25 and 0.125);
\draw[-] (0.525+0.03+0.735,0.05-0.05+2.5) arc (145:35:0.275 and 0.15);

\node at (0.125-0.1+0.5,0+2.5) {$\cdots$};

\draw [decorate,decoration={brace,amplitude=10pt},yshift=4pt,xshift=0pt]
(-1,1.25+2.25) -- (2,1.25+2.25);

\node at (0.5,1.25+3.075) {genus $g_2$};

%%%%%%%%%%%%%%%%%%%%%%%%%%%%%%%%%%%%%%%%%%%%%%%%%%%%%%%%%%%%%%%%%%%%%%%%%

\draw[-] (2.5+1,-0.5) arc (270:450:0.25 and 0.5);
\draw[-] (2.5+1,-0.5) arc (270:90:0.25 and 0.5);

\draw[-] (2.5+1,0.5) .. controls (1.5+1,0.2)  .. (1+1,0.5);
\draw[-] (2.5+1,-0.5) .. controls (1.5+1,-0.2)  .. (1+1,-0.5);

\draw[-] (-1,0.5) .. controls (-1.3,0.2) and (-1.3,-0.2)  .. (-1,-0.5);

\draw[-] (-1,0.5) arc (130:50:1.56+0.775);
\draw[-] (-1,-0.5) arc (-130:-50:1.56+0.775);

\draw[-] (-0.735,0.05) arc (180:360:0.25 and 0.125);
\draw[-] (0.03-0.735,0.05-0.05) arc (145:35:0.275 and 0.15);

\draw[-] (0.525+0.735,0.05) arc (180:360:0.25 and 0.125);
\draw[-] (0.525+0.03+0.735,0.05-0.05) arc (145:35:0.275 and 0.15);

\node at (0.125-0.1+0.5,0) {$\cdots$};

\draw [decorate,decoration={brace,amplitude=10pt},yshift=-4pt,xshift=0pt]
(2,-1) -- (-1,-1);

\node at (0.5,-1.825) {genus $g_1$};

\end{tikzpicture}
}
\caption{Bordisms that generate all physically distinct constructible 2-party states ($N = 2$) in 2D TQFT.}
\label{fig:2partyMaps}
\end{figure}

We now examine the constructible $N$-party states when $n = 2$, focusing on the case $N = 2$ for concreteness. There are two essentially distinct classes of bordisms in $2$\textbf{Cob} for $N = 2$. Under the TQFT functor, they correspond to the connected maps
\begin{equation}
\mathcal{O}_{g}^{0,2} = \delta \circ (\mu \circ \delta)^g \circ \eta
\end{equation}
(see \eqref{genOp}) and the disconnected maps
\begin{equation}
\tilde{\mathcal{O}}_{g_1,g_2}^2 = [(\mu \circ \delta)^{g_1} \circ \eta] \otimes [(\mu\circ\delta)^{g_2} \circ \eta].
\end{equation}
The former has one genus parameter while the latter has two. We depict these maps in Figure \ref{fig:2partyMaps}.

We restrict our attention to TQFTs that can prepare a positive-dimensional submanifold of 1-party states, namely type II TQFTs with $c$ as in \eqref{densec} and $\phi/2\pi\in \mathbb{R}\setminus \mathbb{Q}$. For such TQFTs, the constructible 2-party \textit{disconnected states} are clearly
\begin{equation}
\tilde{\mathcal{O}}^2_{g_1,g_2}(1) \sim (\ket{v_+} - e^{ig_1\phi}\ket{v_-}) \otimes (\ket{v_+} - e^{ig_2\phi}\ket{v_-}).
\end{equation}
These states are separable, and they densely fill a 2-torus consisting of the states
\begin{equation}
\ket{\theta_1,\theta_2}\equiv (\ket{v_+} - e^{i\theta_1} \ket{v_-})\otimes (\ket{v_+} - e^{i\theta_2} \ket{v_-}), \quad \theta_1, \theta_2\in \mathbb{R}.
\end{equation}
On the other hand, by writing the map $\delta$ from Appendix \ref{app:frobenius} in the $\ket{v_+}, \ket{v_-}$ basis, we find that the constructible 2-party \textit{connected states} are
\begin{equation}
\mathcal{O}^{0,2}_{g}(1) \sim \ket{v_+} \otimes \ket{v_+} + e^{i(g+1)\phi}\ket{v_-} \otimes \ket{v_-}.
\end{equation}
These states are entangled, and they densely fill a circle. We see that the constructible 2-party states comprise submanifolds of dimension strictly less than half that of the physical 2-party Hilbert space $\mathbb{CP}^{3}$, in contrast to the 1-party case in which the dimension of the relevant submanifold is precisely half that of the physical Hilbert space.

%%%%%%%%%%%%%%%%%%%%%%%%%%%%%%%%%%%%%%%%%%%%%%
\subsubsection{General Semisimple TQFTs} \label{subsubsec:generalSemisimple}
%%%%%%%%%%%%%%%%%%%%%%%%%%%%%%%%%%%%%%%%%%%%%%

It is easy to see that a generic diagonal handle operator is realizable in a (generally non-unitary) semisimple 2D TQFT, constructed as a direct sum of simple 2D TQFTs with $\dim\mathcal{H}_{S^1} = 1$ \cite{Durhuus:1993cq, Sawin:1995rh}. Indeed, the handle operator of a direct sum of Frobenius algebras $A = A'\oplus A''$ (with counit $\varepsilon = \varepsilon' + \varepsilon''$) is the direct sum of the handle operators of $A'$ and $A''$ \cite{Kock:2004cob}. Taking a direct sum of $n$ 1D Frobenius algebras, we get an $n$-dimensional Frobenius algebra whose handle operator is an arbitrary invertible diagonal matrix.

Specifically, recall from Section \ref{subsubsec:classification} that a simple 2D TQFT $S_z$ is a 1D Frobenius algebra over $\mathbb{C}$ defined as follows. The unit map and multiplication are given by
\begin{equation}
\eta : 1\mapsto 1, \qquad \mu : 1\otimes 1\mapsto 1,
\end{equation}
extending to other elements by linearity. On the other hand, any nonzero linear map $\mathbb{C}\to \mathbb{C}$ defines a counit:
\begin{equation}
\varepsilon : 1\mapsto z^{-1}, \quad z\in \mathbb{C}\setminus \{0\},
\end{equation}
where different values of $z$ correspond to non-isomorphic Frobenius structures. The TQFT axioms then imply that
\begin{equation}
\delta : 1\mapsto z(1\otimes 1).
\end{equation}
Hence the handle operator $\mu\circ \delta$ is multiplication by $z$.

Now consider the semisimple TQFT $\bigoplus_{i=0}^{n-1} S_{z_i}$. Choosing a basis $\{\ket{0}, \ldots, \ket{n - 1}\}$ for $\mathbb{C}^n$, we can write
\begin{equation}
\varepsilon: \ket{i}\mapsto z_i^{-1}, \qquad \mu\circ\delta : \ket{i}\mapsto z_i\ket{i}, \qquad \delta : \ket{i}\mapsto z_i\ket{ii},
\end{equation}
where $\mu : \ket{ij}\mapsto \delta_{ij}\ket{i}$ (no sum) and the multiplicative identity element is $\ket{0} + \cdots + \ket{n - 1}$. This TQFT assigns the invariant
\begin{equation}
Z(\Sigma_g) = \sum_{i=0}^{n-1} z_i^{g-1}
\end{equation}
to a closed, connected surface of genus $g$. For example, for the TQFTs of Section \ref{subsubsec:semisimpleEx}, this invariant takes the form $\sum_{i=0}^{n-1} n^{g-1} = n^g$.

We now return to the questions of universality and complexity in semisimple 2D TQFT. As emphasized in Section \ref{subsec:categoricaldef}, the only fundamental topological gate in the image of the TQFT functor is the handle operator, in the sense that the operations of ``fan-in,'' ``fan-out,'' and crossing don't essentially increase the number of states that we can access in $\mathcal{H}_{S^1}^{\otimes N}$. In particular, the attainable 1-party states are the images of a reference state under nonnegative powers of the handle operator. In a semisimple TQFT, the handle operator rescales the components of a given state in the directions of the eigenstates. In particular, repeated application of this operator magnifies the component in the direction of the eigenvector with largest eigenvalue. If there exist multiple eigenvalues of largest magnitude but with different relative phases, then the image converges to a nontrivial ($\dim > 0$) torus rather than a single accumulation point. States in this torus allow for a one-to-one correspondence between bordism complexity and state complexity. In other words, the TQFT path integral is universal for this set of states.

To be precise, consider a generic semisimple TQFT.  Without loss of generality, we may rescale the handle operator so that it has unit spectral radius (eigenvalues of magnitude $\leq 1$) and order the eigenvalues in decreasing order of magnitude.  The handle operator then becomes
\begin{equation}
H = \operatorname{diag}(e^{i\varphi_0}, \ldots, e^{i\varphi_{m-1}}, r_m e^{i\varphi_m}, \ldots, r_{n-1}e^{i\varphi_{n-1}})
\label{generichandle}
\end{equation}
for some phases $\varphi_0, \ldots, \varphi_{n-1}\in \mathbb{R}$ and magnitudes $r_m, \ldots, r_{n-1}\in \mathbb{R}_{>0}$, where $1 > r_m\geq \cdots\geq r_{n-1}$ and we denote the number of largest eigenvalues by $m$ ($1\leq m\leq n$).  Generically, the phase angles are irrational multiples of $2\pi$ and incommensurate (linearly independent over $\mathbb{Q}$).

Consider a generic reference state $\ket{\psi_0} = (c_0, \ldots, c_{n-1})$.  Let $S$ denote the set of images of this state under powers of the handle operator: $S\equiv \{H^k{\ket{\psi_0}} \,|\, k\geq 0\}$.  From \eqref{generichandle}, any limit point of $S$ must have vanishing amplitude in the $m, \ldots, n - 1$ directions and amplitudes of magnitude $|c_0|, \ldots, |c_{m-1}|$ in the $0, \ldots, m - 1$ directions.  Conversely, assuming that the normalized phases $\hat{\varphi}_0\equiv \varphi_0/2\pi, \ldots, \hat{\varphi}_{m-1}\equiv \varphi_{m-1}/2\pi$ are both irrational and rationally independent, iterating the rotation $(e^{i\varphi_0}, \ldots, e^{i\varphi_{m-1}})$ suffices to approximate any rotation on the $m$-torus to arbitrary accuracy.\footnote{Let $T$ denote the closure of the additive subgroup generated by $(\hat{\varphi}_0, \ldots, \hat{\varphi}_{m-1})$ in $\mathbb{R}^m/\mathbb{Z}^m$ (which may have multiple connected components).  Then we have \cite{24411}:
\begin{equation}
\dim T = \dim_{\mathbb{Q}}\langle 1, \hat{\varphi}_0, \ldots, \hat{\varphi}_{m-1}\rangle - 1.
\end{equation}
In particular, if $\hat{\varphi}_0, \ldots, \hat{\varphi}_{m-1}$ are irrational and rationally independent, then $T = \mathbb{R}^m/\mathbb{Z}^m$.}  Therefore, iterating $H$ suffices to prepare any state on the torus
\begin{equation}
(|c_0|e^{i\theta_0}, \ldots, |c_{m-1}|e^{i\theta_{m-1}}, 0, \ldots, 0), \qquad \theta_0, \ldots, \theta_{m-1}\in \mathbb{R}
\label{thetorus}
\end{equation}
to arbitrary accuracy starting from $\ket{\psi_0}$, as the components in the $m, \ldots, n - 1$ directions can be made arbitrarily small.  We conclude that the limit points of $S$, namely the states that can be approximated to arbitrary accuracy by the TQFT path integral starting from the reference $\ket{\psi_0}$, comprise precisely the torus \eqref{thetorus}.  Every neighborhood of such a state contains infinitely many points of $S$.

This settles the question of universality. One can then ask: how many times must $H$ be iterated to attain a target state to a given accuracy?

Rather than giving a complete answer to this question, we merely note that there exists a simple answer when $n = 2$.\footnote{In the remainder of this section, we use $\mu$ and $\delta$ to denote specific numerical quantities rather than the multiplication and comultiplication operations in a Frobenius algebra. We apologize for the degeneracy of notation.} The starting point is the following basic result on approximating rotations on a circle by powers of a fixed rotation \cite{freedman2021symmetry}. Suppose $\varphi/2\pi$ is irrational with \emph{irrationality measure} $\mu$.\footnote{The irrationality measure $\mu$ of a real number $x$ is defined as the smallest number such that for any $\epsilon > 0$, we have $|x - p/q| > 1/q^{\mu + \epsilon}$ for all integers $p$ and $q$ with $q$ sufficiently large. Almost all real numbers have $\mu = 2$.} Then for any $\epsilon > 0$, for any $\phi$ and any $\delta > 0$, there exists some integer $k$ with $|k\varphi - \phi|\leq \delta$ and with
\begin{equation}
k = O\left(\frac{1}{\delta^{\mu + \epsilon}}\right),
\label{circlebound}
\end{equation}
where the absolute value signs denote distance modulo $2\pi$. The constant factor in these asymptotics depends on $\epsilon$.

Now note that a semisimple 2D TQFT with $n = 2$ that admits some degree of state universality in the sense described above necessarily also has $m = 2$, i.e., both eigenvalues of its handle operator $H$ have equal magnitude. As an operator on the projective Hilbert space $\mathbb{CP}^1$, we may therefore write it as $H = \operatorname{diag}(1, e^{i\varphi})$ for some phase $\varphi$, where we assume $\varphi/2\pi\in \mathbb{R}\setminus \mathbb{Q}$. Then, writing an arbitrary 1-qubit reference state in the Bloch sphere parametrization of $\mathbb{CP}^1$ as $\ket{\theta_0, \phi_0}$ for some $\theta_0\in [0, \pi]$ and $\phi_0\in [0, 2\pi)$, we have
\begin{equation}
H^g\ket{\theta_0, \phi_0} = \cos\left(\frac{\theta_0}{2}\right)\ket{0} + e^{i(\phi_0 + g\varphi)}\sin\left(\frac{\theta_0}{2}\right)\ket{1}.
\end{equation}
Therefore, by choosing $g$ appropriately, we may prepare any state of the form $\ket{\theta_0, \phi}$ for $\phi\in [0, 2\pi)$ to arbitrary accuracy. The result \eqref{circlebound} directly gives an upper bound\footnote{For lower rather than upper bounds on gate complexity, see \cite{bulchandani2021smooth}.} on the complexity of any such state $\ket{\theta_0, \phi}$, with tolerance $\delta$.\footnote{In generalizing this result to the $m$-torus, we would consider a rotation $(e^{i\varphi_0}, \ldots, e^{i\varphi_{m-1}})$ where the normalized angles $\hat{\varphi}_i\equiv \varphi_i/2\pi$ are irrational and rationally independent, so that powers of this rotation are dense in the set of all rotations of the $m$-torus. For a given $\delta$ and a given $(e^{i\phi_0}, \ldots, e^{i\phi_{m-1}})$, we would then seek to bound $k$ such that $|k\varphi_i - \phi_i|\leq \delta$ for all $i$. The answer is not so simple to state because it depends on the ``extent to which'' the $\hat{\varphi}_i$ are linearly independent over $\mathbb{Q}$ (or $\mathbb{Z}$).

On the other hand, rather than bounding the exact complexity, one can argue that the \emph{average-case} complexity is polynomial in $1/\delta$. If all of the rotation angles $\varphi_i$ and target angles $\phi_i$ are chosen uniformly at random from $[0, 2\pi)$, then on average, one would expect to apply the rotation on the order of $1/\delta^m$ times to approximate the target to accuracy $\delta$ as $\delta\to 0$.}

In summary, we have characterized the complexity of semisimple 2D TQFTs in terms of irrational rotations on a torus.

%%%%%%%%%%%%%%%%%%%%%%%%%%%%%%%
\section{Applications and Extensions} \label{sec:applicationsExtensions}
%%%%%%%%%%%%%%%%%%%%%%%%%%%%%%%

One can draw various interesting analogies between 2D TQFTs and other quantum-mechanical systems, such as many-particle systems, tensor networks, and anyons. In this section, we mention some of these analogies, permitting ourselves some room for speculation.

%%%%%%%%%%%%%%%%%%%%%%%%%%%%%%%
\subsection{Enriched TQFT}\label{subsec:enrichedTQFT}
%%%%%%%%%%%%%%%%%%%%%%%%%%%%%%%

We have regarded 2D TQFT as a model for topology change in quantum field theory. These topology-changing operations are \emph{inherently} non-unitary, in contrast to non-unitary operations on open quantum systems.\footnote{In quantum mechanics, non-unitarity arises only from adding or discarding subsystems, i.e., from tensoring with an extra system or taking a partial trace.} It is natural to wonder whether there exists a framework that allows one to formulate the quantum complexity of unitary gates and non-unitary topology-changing processes in a uniform way---e.g., to define a cost function that encompasses both simultaneously.

We do not attempt to answer this question here (one obstacle is that Euclidean and Lorentzian time are not directly comparable). Rather, we simply remark that one language in which to potentially approach this question is that of second quantization in ordinary quantum mechanics.

\begin{figure}
\centering
\begin{tikzpicture}[scale=0.85]
%\draw[-,color=white] (-2.5,0) ellipse (0.25 and 0.5);
%\draw[-,color=white] (2.5,0) ellipse (0.25 and 0.5);

\draw[-,dashed] (-2.5,-0.5) arc (270:450:0.25 and 0.5);
\draw[-] (-2.5,-0.5) arc (270:90:0.25 and 0.5);

\draw[-] (2.5,-0.5) arc (270:450:0.25 and 0.5);
\draw[-] (2.5,-0.5) arc (270:90:0.25 and 0.5);

\draw[-] (-2.5,0.5) .. controls (-1.5,0.2)  .. (-1,0.5);
\draw[-] (2.5,0.5) .. controls (1.5,0.2)  .. (1,0.5);

\draw[-] (-2.5,-0.5) .. controls (-1.5,-0.2)  .. (-1,-0.5);
\draw[-] (2.5,-0.5) .. controls (1.5,-0.2)  .. (1,-0.5);

\draw[-] (-1,0.5) arc (130:50:1.56);
\draw[-] (-1,-0.5) arc (-130:-50:1.56);

\draw[-] (0-0.24,0.1) arc (180:360:0.25 and 0.125);
\draw[-] (0.03-0.24,0.05) arc (145:35:0.275 and 0.15);

\draw[-,color=red,dashed,thick] (0,0.1+0.4375-0.05) ellipse (0.1 and 0.375);
\draw[-,color=blue,dashed,thick] (0,-0.445) ellipse (0.1 and 0.415); 

\node[red] at (0,1.2) {$U_1$};
\node[blue] at (0,-1.2) {$U_2$};
\end{tikzpicture}
\caption{Schematic of a decorated bordism corresponding to evaluation of the path integral in an enriched TQFT. Each colored circle is a separate ``universe'' on which unitary transformations $U_{1,2}: \mathcal{H}_{S^1} \to \mathcal{H}_{S^1}$ may act.}
\label{figs:parentUniv}
\end{figure}
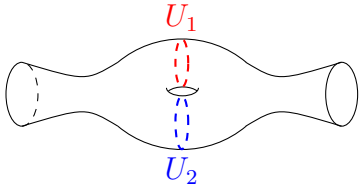

To motivate this picture, note that within a given (Euclidean) time slice, each circle can be thought of as representing a disjoint ``universe,'' and the role of the TQFT dynamics is to produce entanglement between universes. Within an individual universe, it makes sense to consider unitary Lorentzian time evolution according to dynamics that are external to the TQFT. Correspondingly, we define an ``enriched'' 2D TQFT as one whose non-unitary topology-changing operations are supplemented by arbitrary unitary operations on 1-party Hilbert spaces. That is, an \emph{enriched TQFT} is a 2D TQFT in which each tube can be decorated with a unitary operation (Figure \ref{figs:parentUniv}).

We raise two points in relation to this idea:
\begin{enumerate}
\item By introducing an infinite-dimensional ``meta-Hilbert space'' (Fock space) of circles, we obtain a unified framework in which to discuss both standard unitary transformations and topology-changing operations. The latter operations change the occupation number of circles and are therefore represented non-unitarily on the meta-Hilbert space. The problem of defining complexity in enriched TQFT is then mapped to that of defining complexity in second-quantized Fock space.\footnote{We emphasize that, despite superficial similarities between our language and that of ``baby universes'' \cite{Coleman:1988cy, Giddings:1988cx, Giddings:1988wv, Polchinski:1994zs} (see \cite{Marolf:2020xie, McNamara:2020uza, Balasubramanian:2020jhl, Saad:2021uzi} for some recent revivals of this idea, and in particular \cite{Gardiner:2020vjp, deMelloKoch:2021lqp} for connections to 2D TQFT), we are \emph{not} dealing with a theory of dynamical gravity. The spacetime topology is fixed, not summed over. This is merely second quantization, not third quantization.}
\item By construction, an enriched TQFT enjoys 1-party universality (any 1-party state can be constructed from any 1-party reference state) and therefore has access to a larger portion of the $N$-party Hilbert space than an ordinary TQFT. It is natural to ask how much of the $N$-party Hilbert space it can access. In particular, does multipartite universality follow from 1-party universality?
\end{enumerate}
We elaborate further on these two points below.

For clarity, the most concrete calculations below are carried out using qubits, for which $\dim\mathcal{H}_{S^1} = 2$. However, we keep much of the discussion general to any (finite) Hilbert space dimension.

%%%%%%%%%%%%%%%%%%%%%%%%%%%%%%%
\subsubsection{Fock Space of Circles}\label{subsubsec:fockSpaceCircles}
%%%%%%%%%%%%%%%%%%%%%%%%%%%%%%%

We would like to spell out the relation between creation/annihilation of circles and topology change. To do so, we define the ``many-circle Hilbert space'' of the TQFT as the direct sum of all tensor powers of the $S^1$ Hilbert space:
\begin{equation}
\mathcal{H}\equiv \bigoplus_{N=0}^\infty \mathcal{H}_{S^1}^{\otimes N}.
\label{manycircle}
\end{equation}
Implicit in this definition is a notion of ordering of the circles, which allows us to apply TQFT operations accordingly. The symmetric sector of $\mathcal{H}$, consisting of states of the TQFT that are symmetrized over all circles, is isomorphic to a bosonic Fock space:
\begin{equation}
\mathcal{H}\supset \mathcal{H}_B\equiv \bigoplus_{N=0}^\infty \operatorname{Sym}^N(\mathcal{H}_{S^1}).
\label{fockspace}
\end{equation}
The Fock space is constructed as follows. We define operators $a_i, a_i^\dag$ satisfying $[a_i, a_i^\dag] = 1$ ($i = 0, \ldots, n - 1$), where $a_i^\dag$ creates a circle in the state $\ket{i}$ when acting on the Fock vacuum $\ket{\text{vac}}$. The action of $a_i, a_i^\dag$ on the properly normalized Fock states
\begin{equation}
\ket{N_0, \ldots, N_{n-1}}\equiv \frac{(a_0^\dag)^{N_0}}{\sqrt{N_0!}}\cdots \frac{(a_{n-1}^\dag)^{N_{n-1}}}{\sqrt{N_{n-1}!}}\ket{\text{vac}},
\end{equation}
which form a basis (the ``occupation number basis'') for $\mathcal{H}_B$, is as follows:
\begin{align}
a_i\ket{\ldots, N_i, \ldots} &= \sqrt{N_i}\ket{\ldots, N_i - 1, \ldots}, \\
a_i^\dag\ket{\ldots, N_i, \ldots} &= \sqrt{N_i + 1}\ket{\ldots, N_i + 1, \ldots}.
\end{align}
In terms of the standard basis for $\mathcal{H}$ (the ``circle basis''), we have
\begin{equation}
\ket{N_0, \ldots, N_{n-1}} = \frac{1}{\sqrt{N!N_0!\cdots N_{n-1}!}}\sum_{\alpha\in S_N} \alpha{\left(\ket{0^{N_0}\cdots (n - 1)^{N_{n-1}}}\right)},
\label{symmetricsum}
\end{equation}
where $N = N_0 + \cdots + N_{n-1}$ and $S_N$ is the symmetric group on $N$ elements. We have introduced the shorthand notation
\begin{equation}
\ket{0^{N_0}\cdots (n - 1)^{N_{n-1}}}\equiv \ket{0}^{\otimes N_0}\otimes \cdots\otimes \ket{n - 1}^{\otimes N_{n-1}}.
\end{equation}
The sum in \eqref{symmetricsum} contains $N!/(N_0!\cdots N_{n-1}!)$ distinct terms.

Any linear map $T : \mathcal{H}_{S^1}^{\otimes p}\to \mathcal{H}_{S^1}^{\otimes q}$ descends to a linear map $\operatorname{Sym}(T) : \operatorname{Sym}^p(\mathcal{H}_{S^1}) \allowbreak \to \operatorname{Sym}^q(\mathcal{H}_{S^1})$ by symmetrization. Hence the morphisms induced by the TQFT path integral, appropriately symmetrized, can be interpreted as linear operators on the subspace \eqref{fockspace} (note that the swap morphism acts trivially). It is straightforward to write these symmetrized TQFT operations as linear operators on the Fock space \eqref{fockspace} in the occupation number basis. We illustrate this procedure in the case that $n = 2$. The classification of Appendix \ref{app:frobenius} allows us to consider all such two-level TQFTs, which are classified as type I or type II according to \eqref{sigmazero} or \eqref{sigmanonzero}.

For illustration, consider the following four processes that could occur in a symmetric state of $N$ circles: the creation of a single circle, the annihilation of a single circle, the joining of any two circles to form a single one, and the splitting of any circle into two. We denote these processes by the following linear maps from $\operatorname{Sym}^N(\mathcal{H}_{S^1})$ to either $\operatorname{Sym}^{N+1}(\mathcal{H}_{S^1})$ or $\operatorname{Sym}^{N-1}(\mathcal{H}_{S^1})$, which are built on the elementary TQFT operations $\eta$, $\varepsilon$, $\mu$, $\delta$:
\begin{alignat}{2}
T_\text{birth}^N &\equiv \operatorname{Sym}\left(\eta\otimes \operatorname{id}_{S^1}^{\otimes N\vphantom{()}}\right), \qquad & T_\text{join}^N &\equiv \operatorname{Sym}\left(\mu\otimes \operatorname{id}_{S^1}^{\otimes (N-2)}\right), \\
T_\text{death}^N &\equiv \operatorname{Sym}\left(\varepsilon\otimes \operatorname{id}_{S^1}^{\otimes (N-1)}\right), \qquad & T_\text{split}^N &\equiv \operatorname{Sym}\left(\delta\otimes \operatorname{id}_{S^1}^{\otimes (N-1)}\right).
\end{alignat}
To determine the action of these operators on the Fock states of the TQFT, it is convenient to work in the circle basis and to leave symmetrization implicit. Thus we have, for instance (taking $n = 2$),
\begin{equation}
\ket{N_0, N_1} = \sqrt{\frac{N!}{N_0!N_1!}}\ket{0^{N_0}1^{N_1}},
\end{equation}
where the equality is understood in $\operatorname{Sym}^N(\mathcal{H}_{S^1})$ ($N = N_0 + N_1$). The unit $\eta$ simply creates a circle in the $\ket{0}$ state, so that
\begin{equation}
T_\text{birth}^N(\ket{N_0, N_1}) = \eta(1)\otimes \ket{N_0, N_1} = \sqrt{\frac{N!}{N_0!N_1!}}\ket{0^{N_0 + 1}1^{N_1}} = \frac{a_0^\dag}{\sqrt{N + 1}}\ket{N_0, N_1}.
\end{equation}
On the other hand, for any symmetric linear map $T : \mathcal{H}_{S^1}^{\otimes p}\to \mathcal{H}_{S^1}^{\otimes q}$ with $p\leq N$, we can write the action of $T^N\equiv \operatorname{Sym}(T\otimes \operatorname{id}_{S^1}^{\otimes(N - p)})$ as
\begin{equation}
T^N(\ket{N_0, N_1}) = \sqrt{\frac{N_0!N_1!}{N!}}\sum_{k=0}^p \binom{p}{k}\binom{N - p}{N_1 - k}T(\ket{0^{p-k}1^k})\otimes \ket{0^{N_0 + k - p}1^{N_1 - k}},
\end{equation}
with the right side re-symmetrized as necessary. We thus compute, using \eqref{sigmazero} and \eqref{sigmanonzero} for type I and type II TQFTs, that the remaining maps are as follows:
\begin{align}
T_\text{death}^N &= \frac{1}{\sqrt{N}}\times \begin{cases} a_1 & (\text{I}), \\ da_0 & (\text{II}), \end{cases} \\
T_\text{join}^N &= \frac{1}{(N - 1)\sqrt{N}}\times \begin{cases} a_0^\dag a_0^2 + 2a_0 a_1^\dag a_1 + ca_0^\dag a_1^2 & (\text{I}), \\ a_0^\dag a_0^2 + 2a_0 a_1^\dag a_1 + a_0^\dag a_1^2 + ca_1^\dag a_1^2 & (\text{II}), \end{cases} \\
T_\text{split}^N &= \frac{1}{N\sqrt{N + 1}}\times \begin{cases} 2a_0^\dag a_0 a_1^\dag + c(a_0^\dag)^2 a_1 + (a_1^\dag)^2 a_1 & (\text{I}), \\ d^{-1}\left[(a_0^\dag)^2 a_0 + a_0(a_1^\dag)^2 + 2a_0^\dag a_1^\dag a_1 + c(a_1^\dag)^2 a_1\right] & (\text{II}), \end{cases}
\end{align}
where the operators on the left and right of the above equations have identical actions on Fock states $\ket{N_0, N_1}$. Up to a factor that depends on the eigenvalue $N$ of the total circle number operator $a_0^\dag a_0 + a_1^\dag a_1$, all of these TQFT maps can be written as normal-ordered polynomials in the circle creation and annihilation operators.\footnote{More generally, the symmetrization of any linear map $T : \mathcal{H}_{S^1}^{\otimes p}\to \mathcal{H}_{S^1}^{\otimes q}$ given by $T : \ket{i_1\cdots i_p}\mapsto \sum_{j_1, \ldots, j_q} T^{i_1\cdots i_p}{}_{j_1\cdots j_q}\ket{j_1\cdots j_q}$ has the Fock space representation
\begin{equation}
\operatorname{Sym}(T) = \frac{1}{\sqrt{p!q!}}\sum_{i_1, \ldots, i_p}\sum_{j_1, \ldots, j_q} T^{i_1\cdots i_p}{}_{j_1\cdots j_q}(a_{j_1})^\dag\cdots (a_{j_q})^\dag a_{i_1}\cdots a_{i_p}.
\end{equation}} These maps do not preserve the normalization of states.\footnote{An exception to this statement is that in type II TQFTs, the comultiplication operation $\delta$ is an isometry when $c = 0$ and $d = \sqrt{2}$.}

Conversely, it is interesting to note that in type I TQFTs, the bosonic creation and annihilation operators can be written purely in terms of topological maps:
\begin{gather}
a_0^\dag = \sqrt{N + 1}\operatorname{Sym}\left(\eta\otimes \operatorname{id}_{S^1}^{\otimes N\vphantom{()}}\right), \qquad a_0 = \frac{\sqrt{N}}{2}\operatorname{Sym}\left[(\varepsilon\circ\mu\circ\delta)\otimes \operatorname{id}_{S^1}^{\otimes (N-1)}\right], \\
a_1^\dag = \frac{\sqrt{N + 1}}{2}\operatorname{Sym}\left[(\mu\circ\delta\circ\eta)\otimes \operatorname{id}_{S^1}^{\otimes N\vphantom{()}}\right], \qquad a_1 = \sqrt{N}\operatorname{Sym}\left(\varepsilon\otimes \operatorname{id}_{S^1}^{\otimes (N-1)}\right),
\end{gather}
where $a_i^\dag : \operatorname{Sym}^N(\mathcal{H}_{S^1})\to \operatorname{Sym}^{N+1}(\mathcal{H}_{S^1})$ and $a_i : \operatorname{Sym}^N(\mathcal{H}_{S^1})\to \operatorname{Sym}^{N-1}(\mathcal{H}_{S^1})$. The key property in this case is that the handle operator $\mu\circ\delta$ maps the states $\ket{0}$ and $\ket{1}$ into each other: in particular, $(\mu\circ\delta)\ket{0} = 2\ket{1}$.

Finally, we note in passing that a similar story can be developed for the antisymmetric sector of \eqref{manycircle}, namely
\begin{equation}
\mathcal{H}\supset \mathcal{H}_F\equiv \bigoplus_{N=0}^n \bigwedge\nolimits^N(\mathcal{H}_{S^1}),
\end{equation}
which is isomorphic to a fermionic Fock space. In this subspace, the swap morphism effects a sign flip. However, the fermionic story is far less interesting than the bosonic one in our context, both due to the finite number of nonzero Fock states and because (co)commutativity implies that the images of $\mu$ and $\delta$ are trivial.

%%%%%%%%%%%%%%%%%%%%%%%%%%%%%%%
\subsubsection{Toward Universality}\label{subsubsec:blochSphere}
%%%%%%%%%%%%%%%%%%%%%%%%%%%%%%%

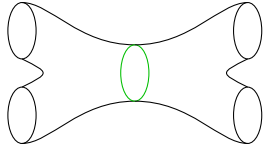
\begin{figure}
\centering
\begin{tikzpicture}[scale=0.75]
%\draw[-,white] (0,1.5) ellipse (0.25 and 0.5);
%\draw[-,white] (0,-1.5) ellipse (0.25 and 0.5);

\draw[-] (-2,0.75) ellipse (0.25 and 0.5);
\draw[-] (-2,-0.75) ellipse (0.25 and 0.5);

\draw[-] (0,0.5) .. controls (-1,0.5) and (-1.5,1.25) .. (-2,1.25);
\draw[-] (0,-0.5) .. controls (-1,-0.5) and (-1.5,-1.25) .. (-2,-1.25);

\draw[-] (-2,0.25) .. controls (-1.5,0) .. (-2,-0.25);
\draw[-,black!25!green] (0,0) ellipse (0.25 and 0.5);

\draw[-] (0,0.5) .. controls (1,0.5) and (1.5,1.25) .. (2,1.25);
\draw[-] (0,-0.5) .. controls (1,-0.5) and (1.5,-1.25) .. (2,-1.25);

\draw[-] (2,0.25) .. controls (1.5,0) .. (2,-0.25);

\draw[-] (2,0.75) ellipse (0.25 and 0.5);
\draw[-] (2,-0.75) ellipse (0.25 and 0.5);

%\node at (-1,0) {$\mu$};
%\node at (1,0) {$\delta$};

\end{tikzpicture}
\caption{The composition $\delta \circ \mu$ of comultiplication and multiplication. This operator creates entanglement between the two input universes. The same is true of decorated versions of this bordism.}
\label{fig:entanglementCreator}
\end{figure}

A more immediate question than that of complexity is whether enriched TQFTs, defined as 2D TQFTs in which arbitrary local (1-party) unitaries are allowed to act on individual circles,\footnote{In fact, it suffices to consider the action of two judiciously chosen unitaries because two generic unitaries generate a dense subset of $SU(n)$, the group of unitaries on the projective Hilbert space $\mathbb{CP}^{n-1}$ of an $n$-level qudit \cite{kuranishi_1951, Sawicki_2017}. For instance, the Hadamard gate and an appropriately chosen phase gate generate $SU(2)$. An algorithm for finding such unitaries is given in \cite{Sawicki_20172}.} are universal: namely, can they prepare all states in their $N$-party Hilbert space? A first guess might be that the answer is ``yes,'' in light of the following theorems about quantum circuits:
\begin{itemize}
\item A universal 1-qudit gate set plus any 2-qudit gate that does not map separable states to separable states is universal for quantum computation \cite{2001quant.ph..8062B}. In other words, ``1-qudit universality $+$ 2-qudit entanglement $=$ $N$-qudit universality.''
\item 2-qubit gates are exactly universal for quantum computation \cite{preskillnotes}.
\end{itemize}
Unfortunately, neither theorem applies in our context because the TQFT analogues of 2-party ``gates'' (linear maps from $\mathcal{H}_{S^1}^{\otimes 2}\to \mathcal{H}_{S^1}^{\otimes 2}$) are weaker than unitaries by virtue of being non-invertible. For instance, the operation $\delta\circ\mu$ of multiplication followed by comultiplication creates entanglement when acting on separable 2-party states (Figure \ref{fig:entanglementCreator}), but this map has rank $\dim\mathcal{H}_{S^1} < (\dim\mathcal{H}_{S^1})^2$ and is therefore not invertible. This is simply because the image of $\delta$ has dimension $\dim\mathcal{H}_{S^1}$ inside $\mathcal{H}_{S^1}^{\otimes 2}$.

In fact, by considering the normal form, we see that any linear map induced by a connected bordism from $p$ to $q$ circles has rank at most $\dim\mathcal{H}_{S^1}$ (the dimensional ``bottleneck'' being the cylinder), and this remains true for enriched TQFT. Due to these rank restrictions, it is impossible to approximate any 2-party unitary to arbitrary accuracy using TQFT operations combined with 1-party unitaries.\footnote{A universal gate set for non-unitary quantum circuits is exhibited in \cite{nonunitary}, but those results do not apply here because they rely on the unitary CNOT gate.}

Rather than addressing the question of whether enriched TQFTs are universal in complete generality, we make two simplifications. First, as a small step toward answering this question, we ask whether enriched TQFTs can prepare all 2-party states (rather than $N$-party states for arbitrary $N$). Second, we focus on the case of qubits ($n = 2$). In fact, we further restrict to 2D TQFTs of type I, for which the subsequent calculations are especially simple.

To begin, note that by the assumption of 1-party universality, an enriched TQFT can prepare any separable state $\ket{\psi_1}\otimes \ket{\psi_2}\in \mathcal{H}_{S^1}^{\otimes 2}$. We would like to know whether, by applying a TQFT morphism that acts as an entangler (the simplest of which is $\delta\circ\mu$) in addition to local unitaries (which preserve the entanglement structure of a bipartite state), one can obtain any $\ket{\psi}\in \mathcal{H}_{S^1}^{\otimes 2}$. In fact, because any entangler such as $\delta\circ\mu$ can be written as the composition of a morphism $\mathcal{H}_{S^1}\to \mathcal{H}_{S^1}^{\otimes 2}$ with a morphism $\mathcal{H}_{S^1}^{\otimes 2}\to \mathcal{H}_{S^1}$, and universality on $\mathcal{H}_{S^1}$ is assumed, the question of 2-party universality in enriched TQFT boils down to:
\begin{quote}
\emph{Can the comultiplication operation $\delta$ create all possible kinds of entanglement between two qubits, starting from an arbitrary 1-qubit state?}
\end{quote}
More precisely, two pure states of a bipartite quantum system are locally unitarily equivalent if and only if their Schmidt coefficients (or alternatively, reduced density matrix eigenvalues) coincide. To show that an enriched TQFT is universal on two parties, it suffices to show that the image of $\delta$ contains states with all possible Schmidt coefficients.

To this end, we consider an arbitrary type I TQFT specified by a single complex number $c$, with the corresponding map $\delta$ given in \eqref{sigmazero}. An arbitrary 1-qubit state $\cos(\theta/2)\ket{0} + e^{i\phi}\sin(\theta/2)\ket{1}$ with $\theta\in [0, \pi]$ and $\phi\in [0, 2\pi)$ maps via $\delta$ to the 2-qubit state
\begin{equation}
\ket{\psi_{12}}\propto e^{i\phi}\sin\left(\frac{\theta}{2}\right)(c\ket{00} + \ket{11}) + \cos\left(\frac{\theta}{2}\right)(\ket{01} + \ket{10}),
\end{equation}
up to normalization, whose reduced density matrix is
\begin{equation}
\rho_1 = \Tr_2\dyad{\psi_{12}}\propto \begin{pmatrix}
\cos^2\left(\dfrac{\theta}{2}\right) + |c|^2\sin^2\left(\dfrac{\theta}{2}\right) &\ \ & \dfrac{1}{2}\sin(\theta)(e^{-i\phi} + ce^{i\phi})\\
\dfrac{1}{2}\sin(\theta)(e^{i\phi} + c^\ast e^{-i\phi}) &\ \ & 1
\end{pmatrix}.
\end{equation}
The state $\ket{\psi_{12}}$ is maximally mixed when $\theta = 0$, regardless of $\phi$ and $c$ (in which case the largest eigenvalue of $\rho_1$, properly normalized, is $1/2$). Since the largest eigenvalue of $\rho_1$ is a continuous function of $(\theta, \phi)$, to show that it attains all possible values in the range $[1/2, 1]$ as $(\theta, \phi)$ range over the entire Bloch sphere, it suffices to show that for any given $c$, there exist some $(\theta, \phi)$ for which the state $\ket{\psi_{12}}$ is separable (i.e., for which the largest eigenvalue of $\rho_1$ is 1).

The separability of $\ket{\psi_{12}}$ is equivalent to the condition that $\det\rho_1 = 0$, which can be written as
\begin{equation}
\left[x - c'(1 - x)\right]\left[x - c'^\ast(1 - x)\right] = 0, \qquad x\equiv \cos^2\left(\frac{\theta}{2}\right), \qquad c'\equiv ce^{2i\phi}.
\end{equation}
This equation is satisfied when
\begin{equation}
x = \frac{c'}{1 + c'} \quad \text{or} \quad x = \frac{c'^\ast}{1 + c'^\ast}.
\end{equation}
Since $0\leq x\leq 1$, a solution exists only when $c'$ is real and $c'\geq 0$. But by choosing $\phi$, we can satisfy these conditions on $c'$ for any given $c$. Hence there indeed exists a 1-qubit state whose image under $\delta$ is separable, regardless of $c$.

We have thus demonstrated 2-party universality for type I TQFTs with $n = 2$. We leave open the question of whether $N$-party universality follows from 2-party universality. Again, we stress that such a statement does not follow from the universality of 2-qubit quantum gates, because our 2-qubit topological ``gates'' are non-unitary.

%%%%%%%%%%%%%%%%%%%%%%%%%%%%%%%%%%%%
\subsection{Extended TQFT}\label{subsec:extendedTQFT}
%%%%%%%%%%%%%%%%%%%%%%%%%%%%%%%%%%%%

A different way of extending our formalism is to enlarge our conception of functorial QFT, which we have so far used only in the most limited of circumstances. In particular, we have seen that for 2D TQFTs described by commutative Frobenius algebras, the path integral lacks sufficient power to prepare most states in the Hilbert space on a closed 1-manifold. Said differently, our topological ``circuits'' are highly non-universal. This crude axiomatization also precludes the possibility of computing nontrivial \emph{correlation functions} (e.g., of Wilson lines), an option that would be available in TQFTs arising from gauge theories with a Lagrangian description. Such operator insertions would dramatically increase the number of states that one could prepare via the path integral.

One way of going beyond our simple axiomatization is to work in the framework of \emph{extended} QFT, which formalizes in terms of higher categories of bordisms the notion of a fully local QFT (such as one arising naturally in physics) as assigning data to manifolds of arbitrary codimension.\footnote{One could also contemplate simpler modifications of the input category, such as symmetric $\to$ braided rather than ordinary $\to$ higher. A symmetric monoidal category is a special case of a \textit{braided} monoidal category in which the \textit{braid map} squares to the identity. Relaxing this requirement suggests a generalization of 2D TQFT to use braids rather than swaps \cite{Kock:2004cob}. In this ``anyonic'' generalization, the path integral on surfaces is sensitive to their embedding in ambient $\mathbb{R}^3$.} The most prominent examples are extended TQFT \cite{Baez_1995} and (2D) CFT \cite{Runkel:2005qw, henriques2013threetier}, with Frobenius algebras again playing a role in the latter. An advantage of non-topological functorial QFTs is that Euclidean propagation along a cylinder (i.e., in the absence of topology change) is nontrivial, which suggests a way to achieve universality without leaving the setting of Euclidean time (in contrast to the ad hoc construction of Section \ref{subsec:enrichedTQFT}).

Another way of augmenting the computational power of the path integral is to pass directly to theories with Lagrangian descriptions---for instance, 2D BF or Yang-Mills theory, the latter of which is only quasi-topological \cite{Witten:1991we, Cordes:1994fc}. The first step would be to classify the states that can be prepared by the path integral in these cases. A classification of this nature already exists for $U(1)$ and $SO(3)$ Chern-Simons theory on 3-manifolds with torus boundaries \cite{Salton:2016qpp}, where the incorporation of Wilson lines is crucial for achieving state universality in certain cases. Related studies of topological entanglement entropy have also been carried out in Chern-Simons theory \cite{Balasubramanian:2016sro, Balasubramanian:2018por, Fliss:2020yrd, Leigh:2021trp} and in 2D TQFT \cite{Melnikov:2018zfn}.

%%%%%%%%%%%%%%%%%%%%%%%%%%%%%%%%%%%%
\subsection{Holographic Complexity}\label{subsec:holoComplexity}
%%%%%%%%%%%%%%%%%%%%%%%%%%%%%%%%%%%%

So far, our discussion has been restricted to exploring circuit complexity in the context of 2D TQFT. However, a version of the same functorial story also applies to CFT, and this naturally leads one to consider whether this story could tell us anything about holographic complexity, particularly in AdS$_3$/CFT$_2$.

The complexity of a holographic CFT state, modulo the usual ambiguities in definition, has been proposed to be dual to either the volume of a maximal slice in the bulk (``Complexity = Volume,'' or CV) or the action evaluated on the causal development of such a slice (``Complexity = Action,'' or CA) \cite{Susskind:2014rva, Stanford:2014jda, Susskind:2014moa, Brown:2015bva, Brown:2015lvg}. Let us consider an instructive and motivating example: the $n$-sided, genus-$g$ AdS$_3$ wormhole solution. For such geometries, the ``complexity of formation'' according to both CV and CA was computed in \cite{Fu:2018kcp}, with the reference state being $n$ copies of the $M = 0$ BTZ black hole.\footnote{The entanglement structure of the dual CFT states has been studied in \cite{Balasubramanian:2014hda, Marolf:2015vma, Maxfield:2016mwh, Peach:2017npp, Balasubramanian:2018hsu}.} After subtracting the complexity of the reference state, the resulting quantity $\Delta \mathcal{C}$ has the property of being UV-finite. The results of \cite{Fu:2018kcp} are as follows:
\begin{equation}
\Delta\mathcal{C}(n,g) = \begin{cases}
\dfrac{2\pi L}{G}(2g-2+n) = -\dfrac{4}{3}\pi c\chi & \text{for CV},\\[10 pt]
\dfrac{L}{4G}(2-2g-n) = \dfrac{1}{6}c\chi & \text{for CA},
\end{cases}
\end{equation}
where $L$ is the AdS curvature scale, $G$ is Newton's constant, $c = \frac{3L}{2G}$ is the central charge of the CFT, and $\chi = 2 - 2g - n$ is the Euler characteristic.

It is interesting, though perhaps not unexpected, that both of these expressions are proportional to the Euler characteristic of the $t=0$ spatial slice, which in turn measures the size of its pants decomposition. It is tempting to think of this spatial slice as a bordism, and the complexity of formation as quantified by the pants decomposition. A toy model in which this point of view could potentially be justified is the Chern-Simons/chiral Wess-Zumino-Witten duality, which comes in both an ``AdS/CFT-like'' version and a ``dS/CFT-like'' version. The latter of these relates the wavefunction of the CS theory on a spatial slice to a Euclidean WZW theory living on that slice. By viewing this slice as a bordism that prepares a state in the Lorentzian dual theory, we see that such a boundary state can be prepared by either a Lorentzian circuit on the boundary or a Euclidean circuit in the bulk. The complexity of the former may lead to a derivation of the complexity of the latter as a circuit whose gates are pairs of pants.

Finally, note that a 2-bordism, whether representing a spatial slice in a 3D gravitational bulk or the Euclidean spacetime of a 2D field theory, resembles a circuit that prepares a tensor network state from rank-3 tensors (in which each pair of pants is a gate). In a topological theory, the path integral maps non-uniquely onto such a tensor network, or trivalent graph. The tensor network complexity (minimum number of elementary tensors) then coincides with the circuit complexity of the bordism. In our context, the circuit complexity can also be interpreted as a path integral complexity in the sense of \cite{Miyaji:2016mxg, Caputa:2017urj, Caputa:2017yrh}, but one which involves optimizing over topologies rather than over metrics (with respect to a complexity measure that is monotonic in the genus).

%%%%%%%%%%%%%%%%%%%%%%%%%%%%%%%
\section{Conclusion}\label{sec:conclusion}
%%%%%%%%%%%%%%%%%%%%%%%%%%%%%%%

We have argued that the pants decomposition provides a natural notion of complexity in the category $2$\textbf{Cob}, and furthermore, that this induces a notion of complexity on 2D TQFT states prepared by the Euclidean path integral. However, this notion of complexity is far from universal. One avenue for rectifying this deficiency might be to consider TQFT extended down to points, while another might be to include gates on the Hilbert space associated to a single circle which are not represented by bordisms. In addition, we have briefly speculated on applications to AdS$_3$ holography, and in particular to the holographic complexity conjectures.

While this work has focused on 2D TQFT, our hope is that similar considerations apply to higher-dimensional TQFT as well as to CFT. For $d \geq 3$, $d\textbf{Cob}$ has no simple analogue of the pants decomposition, and the gate set relevant for higher-dimensional bordisms would likely need to be infinite, although the language of higher categories may make the generalization more tractable \cite{Carqueville:2017fmn}.\footnote{Modifying the input category or passing to higher dimensions could also allow for heterogeneous gate sets and hence the possibility of different penalty factors for different gates---unlike in 2D, where all elementary gates are essentially the same (pairs of pants).} The same caveats apply to higher-dimensional CFT. An extension of this work to 2D CFT (as required for the proposed holographic applications) may be easier, but is left to future work.

It is worth noting that complexity is not the only information-theoretic quantity that can be studied in the framework of category theory. The axiomatic point of view has also led to quantitative results for entanglement entropy in 2D open-closed TQFT \cite{Donnelly:2018ppr, Lauda:2005wn}. One can contemplate similar generalizations to CFT and to higher dimensions in that context.

We end by pointing out some even more speculative directions for future work:
\begin{itemize}
\item One can extend this story by forgetting structure (e.g., orientation) or by adding structure (e.g., spin structure), or more generally, by extending the analysis to $G$-equivariant TQFTs \cite{deMelloKoch:2021lqp}.

\item Our notion of complexity is strictly discrete. Can one make contact with Nielsen's complexity geometry \cite{Nielsen:2005cg, Nielsen_20061, Nielsen:2006cg} (for a manifold of non-unitary transformations) by combining topological gates and unitary gates? Note that topology change can be used as a mechanism for achieving universality in certain non-universal schemes for topological quantum computation \cite{Salton:2016qpp}.

\item Complexity has been proposed as a general tool in category theory, where it generalizes classical circuit complexity \cite{BASU_2020}. Independently of TQFT, one could ask: how does one ``quantize'' the formalism of \cite{BASU_2020} (i.e., formulate the analogue for quantum circuit complexity)?

\item This work suggests exploring quantum complexity theory in imaginary time.
\end{itemize}

\section*{Acknowledgements}

We thank Scott Aaronson, Elena C\'aceres, Jacques Distler, Willy Fischler, Dan Freed, and Hao-Yu Sun for useful discussions, as well as Shira Chapman, Jacques Distler, and Brian Swingle for valuable comments on a draft. YF thanks the UT quantum information group for inspiration. This work was supported by the National Science Foundation under Grants No.\ PHY-1620610 (JC), PHY-1820712 (JC and SS), and PHY-1914679 (YF and SS).

\appendix

%%%%%%%%%%%%%%%%%%%%%%%%%%%%%%%%%%%%%%%%%%%%%%%%%
\section{Category Theory} \label{app:category}
%%%%%%%%%%%%%%%%%%%%%%%%%%%%%%%%%%%%%%%%%%%%%%%%%

We review here some essential terminology from category theory, ignoring many subtleties. We then introduce the categories $d\textbf{Cob}$ and $\textbf{Hilb}$, which allow us to define an axiomatic TQFT in functorial language.

\begin{figure}
\centering
\begin{tikzpicture}[scale=0.7]
\node[color=red] (1) at (-2,1) {$\bullet$};
\node[color=red] (2) at (-1,2) {$\bullet$};
\node[color=red] (3) at (1,2) {$\bullet$};
\node[color=red] (4) at (2,1) {$\bullet$};
\node[color=red] (5) at (2,-1) {$\bullet$};
\node[color=red] (6) at (1,-2) {$\bullet$};
\node[color=red] (7) at (-1,-2) {$\bullet$};
\node[color=red] (8) at (-2,-1) {$\bullet$};

\draw[->,thick] (1) to (2);
\draw[->,thick] (2) to (3);
\draw[->,thick] (3) to (4);
\draw[->,thick] (4) to (5);
\draw[->,thick] (5) to (6);
\draw[->,thick] (6) to (7);
\draw[->,thick] (7) to (8);
\draw[->,thick] (8) to (1);

\draw[->,thick] (8) to (3);
\draw[->,thick] (3) to (7);
\draw[->,thick] (5) to (3);
\draw[->,thick] (8) to (2);

\draw[->,thick] (1) .. controls (-3+0.5,2+0.5) and (-3-0.5,2-0.5) .. (1);
\draw[->,thick] (2) .. controls (-2+0.5,3+0.5) and (-2-0.5,3-0.5) .. (2);
\draw[->,thick] (3) .. controls (2+0.5,3-0.5) and (2-0.5,3+0.5) .. (3);
\draw[->,thick] (4) .. controls (3+0.5,2-0.5) and (3-0.5,2+0.5) .. (4);
\draw[->,thick] (5) .. controls (3+0.5,-2+0.5) and (3-0.5,-2-0.5) .. (5);
\draw[->,thick] (6) .. controls (2+0.5,-3+0.5) and (2-0.5,-3-0.5) .. (6);
\draw[->,thick] (7) .. controls (-2-0.5,-3+0.5) and (-2+0.5,-3-0.5) .. (7);
\draw[->,thick] (8) .. controls (-3+0.5,-2-0.5) and (-3-0.5,-2+0.5) .. (8);
\end{tikzpicture}
\caption{A category consisting of points and arrows. The loops are identity morphisms, and the composition of any combination of arrows is found by following them in order.}
\label{fig:catDef}
\end{figure}
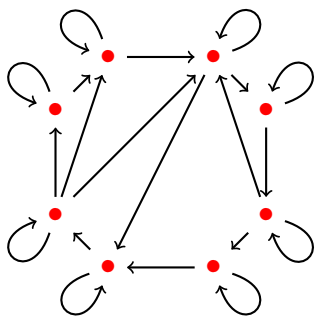

A \textit{category} is a class of \textit{objects} equipped with a class of \textit{morphisms}. The morphisms satisfy three key properties:
\begin{itemize}
\item[(1)] \textit{composition}: for any three objects $a,b,c$ and any two morphisms $f: a \to b$ and $g: b \to c$, there exists a morphism $g \circ f: a \to c$,

\item[(2)] \textit{associativity}: for any three morphisms $f: a \to b$, $g: b \to c$, and $h: c \to d$, the compositions $(h \circ g) \circ f$ and $h \circ (g \circ f)$ are equal,

\item[(3)] \textit{identity}: for any object $a$, there exists a morphism $\text{id}_a: a \to a$ such that for any morphisms $f: a \to b$ and $g: c \to a$, we have $f \circ \text{id}_a = f$ and $\text{id}_a \circ g = g$.
\end{itemize}
For concreteness, we may depict the objects as points and the morphisms as arrows. See Figure \ref{fig:catDef} for an example.

Many categories of interest are infinite. Familiar examples include \textbf{Grp} (the category of groups, with morphisms being group homomorphisms), \textbf{Top} (the category of topological spaces, with morphisms being continuous maps), and \textbf{Vect}$_k$ (the category of vector spaces over a field $k$, with morphisms being linear maps).

Given two categories $\mathcal{C}$ and $\mathcal{D}$, a \emph{functor} is a map $F : \mathcal{C} \to \mathcal{D}$ that sends each object in $\mathcal{C}$ to an object in $\mathcal{D}$ while preserving the morphism structure.  Namely, for any morphism $f: a \to b$ in $\mathcal{C}$, there exists a morphism $F(f): F(a) \to F(b)$ in $\mathcal{D}$ such that $F(\text{id}_a) = \text{id}_{F(a)}$ and $F(g \circ f) = F(g) \circ F(f)$.

A \textit{monoidal category} is a category equipped with a notion of multiplication and a corresponding neutral object, while a \textit{monoidal functor} between such categories preserves the monoidal structure. For example, the category \textbf{Set} is a monoidal category with respect to disjoint union $\sqcup$ (with neutral object $\varnothing$), and the ``swap'' map under which the two sets $A\sqcup B$ and $B\sqcup A$ are canonically isomorphic is the prototype of a \textit{symmetric structure} on such a category.\footnote{Swap morphisms on monoidal categories are examples of \textit{braid morphisms}, where the latter do not necessarily square to the identity. For the categories of interest to us, only swaps (symmetric braids) are needed.} In general, we call $\mathcal{C}$ a \textit{symmetric monoidal category} if there exists a \textit{bifunctor}
\begin{equation}
\otimes: \mathcal{C} \times \mathcal{C} \to \mathcal{C}
\end{equation}
that is associative, commutative, and has an identity element in $\mathcal{C}$, i.e., for any objects $a,b,c \in \mathcal{C}$ and some $1 \in \mathcal{C}$,
\begin{equation}
(a \otimes b) \otimes c \cong a \otimes (b \otimes c), \qquad a \otimes b \cong b \otimes a, \qquad 1 \otimes a \cong a \otimes 1 \cong a,
\end{equation}
where $\cong$ denotes isomorphism. An example is the category \textbf{Vect}$_k$, with the bifunctor being the tensor product and the identity being $k$.

For any category $\mathcal{C}$, we define its \emph{dual} category $\mathcal{C}^{*}$ by reversing the arrows. We call $\mathcal{C}$ a \textit{dagger} or \textit{$\dagger$-category} \cite{Baez:2004pa} if there exists a functor
\begin{equation}
\dagger: \mathcal{C}^* \to \mathcal{C}
\end{equation}
that sends objects to themselves and, for any morphism $f^*: a \to b$ in $\mathcal{C}^*$, sends $f^*$ to a morphism $f^\dagger: a \to b$ in $\mathcal{C}$ such that
\begin{equation}
\text{id}_a^\dagger = \text{id}_a, \qquad (g\circ f)^\dagger = f^\dagger \circ g^\dagger, \qquad (f^\dagger)^\dagger = f. \label{dag}
\end{equation}
This is the category-theoretic version of the Hermitian conjugate, with $f^\dagger$ being called the \textit{adjoint} of $f: b \to a$.

We will find that both $d\textbf{Cob}$ and \textbf{Hilb} may be given a symmetric monoidal structure and a $\dagger$ functor. The TQFT functor always preserves the former property and may also preserve the latter.

%%%%%%%%%%%%%%%%%%%%%%%%%%%%%%%%%%%%
\subsection{Category of Bordisms}\label{subapp:bordisms}
%%%%%%%%%%%%%%%%%%%%%%%%%%%%%%%%%%%%

We define the category $d$\textbf{Cob} for integer $d\geq 1$ as follows: the objects are closed, oriented $(d - 1)$-manifolds, the morphisms are equivalence classes of oriented $d$-dimensional bordisms (bordism classes),\footnote{Formally, for two closed $(d-1)$-manifolds $M$ and $N$, a $d$-dimensional bordism from $M$ to $N$ is a quintuple $(W, M, N, i_M, i_N)$ where $W$ is a compact $d$-manifold with boundary $\partial W$ and $i_M: M \to \partial W$ and $i_N: N \to \partial W$ are embeddings such that
\begin{equation}
i_M(M) \cap i_N(N) = \varnothing, \qquad i_M(M) \cup i_N(N) = \partial W.
\end{equation}
It is a common abuse of terminology simply to refer to the manifold $W$ as such \cite{Freed:2012bon}.} and composition is ``gluing together'' of bordisms.

$(d\textbf{Cob}, \sqcup, \varnothing)$ is a symmetric monoidal category. The relevant bifunctor is disjoint union $\sqcup$, and the neutral object is the empty set $\varnothing$ as a $(d - 1)$-manifold. The symmetric structure is given by the \textit{swap bordism}. We may also give \textbf{$d$Cob} the structure of a $\dagger$-category, where taking the adjoint of a bordism switches the ``input'' and ``output'' $(d-1)$-manifolds.

\begin{figure}
\centering
\begin{tikzpicture}[scale=0.85]
\draw[-,color=white] (-2.5,0) ellipse (0.25 and 0.5);
\node at (-3.2,0) {$M$};

\draw[-,color=white] (2.5,0) ellipse (0.25 and 0.5);
\node at (3.2,0) {$N$};

\draw[-,dashed] (-2.5,-0.5) arc (270:450:0.25 and 0.5);
\draw[-] (-2.5,-0.5) arc (270:90:0.25 and 0.5);

\draw[-] (2.5,-0.5) arc (270:450:0.25 and 0.5);
\draw[-] (2.5,-0.5) arc (270:90:0.25 and 0.5);

\draw[-] (-2.5,0.5) .. controls (-1.5,0.2)  .. (-1,0.5);
\draw[-] (2.5,0.5) .. controls (1.5,0.2)  .. (1,0.5);

\draw[-] (-2.5,-0.5) .. controls (-1.5,-0.2)  .. (-1,-0.5);
\draw[-] (2.5,-0.5) .. controls (1.5,-0.2)  .. (1,-0.5);

\draw[-] (-1,0.5) arc (130:50:1.56);
\draw[-] (-1,-0.5) arc (-130:-50:1.56);

\draw[-] (0-0.24,0.1) arc (180:360:0.25 and 0.125);
\draw[-] (0.03-0.24,0.05) arc (145:35:0.275 and 0.15);

\draw[-,dashed,thick] (0,0.1+0.4375-0.05) ellipse (0.1 and 0.375);
\draw[-,dashed,thick] (0,-0.445) ellipse (0.1 and 0.415); 
\end{tikzpicture}
\caption{A 2-bordism from $M \cong S^1$ to $N \cong S^1$ with genus $g = 1$. Cutting along the dashed lines gives a pants decomposition.}
\label{fig:2cob}
\end{figure}
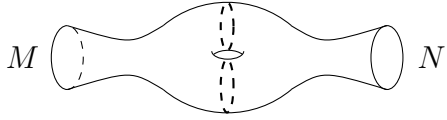

Of particular importance to our work is $2\textbf{Cob}$, discussed extensively in \cite{Kock:2004cob}. The only closed, connected 1-manifold is $S^1$, so the objects are finite disjoint unions of circles. The bordisms are oriented surfaces with arbitrary genus and circle boundaries. An example is shown in Figure \ref{fig:2cob}. Any 2-bordism may be written as a composition and disjoint union of the elementary bordisms shown in Figure \ref{figs:decomposition} \cite{Kock:2004cob}.

%%%%%%%%%%%%%%%%%%%%%%%%%%%%%%%%%%%%
\subsection{Category of Hilbert Spaces}\label{subapp:hilbert}
%%%%%%%%%%%%%%%%%%%%%%%%%%%%%%%%%%%%

We denote by \textbf{Vect}$_{\mathbb{C}}$ the category whose objects are finite-dimensional complex vector spaces and whose morphisms are linear maps. $(\textbf{Vect}_{\mathbb{C}}, \otimes, \mathbb{C})$ is a symmetric monoidal category. The associated bifunctor is the tensor product $\otimes$, and the neutral object is $\mathbb{C}$. The symmetric structure is given by the \textit{swap map} that exchanges the factors of a tensor product.

By equipping \textbf{Vect}$_{\mathbb{C}}$ with a $\dag$-structure, we promote it to the category of finite-dimensional Hilbert spaces \textbf{Hilb} \cite{Baez:2004pa}. By the one-to-one correspondence between vectors in a complex vector space $V$ and linear maps $\mathbb{C}\to V$, this extra structure automatically provides the notions of bra, ket, and conjugate-symmetric inner product $\bra{\psi}\ket{\phi} = \overline{\bra{\phi}\ket{\psi}}$ on all objects in the category.

Note that only the vector space structure of the objects is preserved by generic morphisms. Morphisms that preserve the inner product are called \textit{isometries}, with any unitary operator $U$ being an example: $\bra{\psi}U^\dagger U\ket{\phi} = \bra{\psi}\ket{\phi}$. We note also that the $\dagger$ functor interacts with the $\otimes$ bifunctor as follows: $(f \otimes g)^\dagger = f^\dagger \otimes g^\dagger$.

%%%%%%%%%%%%%%%%%
\subsection{Atiyah's TQFT Functor}\label{subapp:atiyah}
%%%%%%%%%%%%%%%%%

We define a \textit{$d$-dimensional TQFT} \cite{Atiyah:1989vu} as a \textit{symmetric monoidal functor}
\begin{equation}
Z: \textbf{$d$Cob} \to \textbf{Vect}_{\mathbb{C}}.
\end{equation}
Less succinctly, but in more physically intuitive language, the map $Z$ satisfies six axioms, which we state in words (for careful definitions, see \cite{Kock:2004cob}):
\begin{enumerate}
\item Equivalent bordisms have the same image. (Two bordisms are equivalent if they are related by a boundary- and orientation-preserving diffeomorphism.)

\item The cylinder over $\Sigma$ goes to the identity map on $Z(\Sigma)$.

\item Composition of bordisms goes to composition of linear maps.

\item Disjoint union of manifolds (resp.\ bordisms) goes to tensor product of vector spaces (resp.\ linear maps).

\item The empty manifold (resp.\ bordism) goes to the ground field $\mathbb{C}$ (resp.\ the identity map on $\mathbb{C}$).

\item The bordism that interchanges $\Sigma$ and $\Sigma'$ goes to the linear isomorphism $Z(\Sigma)\otimes Z(\Sigma')\to Z(\Sigma')\otimes Z(\Sigma)$ given by $a\otimes a'\mapsto a'\otimes a$.
\end{enumerate}
Axiom 1 says that the map $Z : d\textbf{Cob} \to \textbf{Vect}_{\mathbb{C}}$ is well-defined (depends only on the equivalence class of $M$). Axioms 2 and 3 say that $Z$ is a functor (preserves identity and composition). Axioms 4 and 5 say that this functor preserves the monoidal structure. Axiom 6 says that this functor preserves the symmetric structure. The axioms together imply that for any $\Sigma$, there exists a nondegenerate pairing $Z(\Sigma)\times Z(\overline{\Sigma})\to \mathbb{C}$, which in turn implies that $Z(\Sigma)$ is finite-dimensional \cite{Kock:2004cob}.

The $d$-dimensional TQFT defined by $Z$ is said to be \textit{unitary} if, viewed as a functor from the $\dag$-category $d\textbf{Cob}$ to \textbf{Hilb}, it preserves the $\dagger$-structure \cite{Baez:2004pa}: $Z(M^\dagger) = Z(M)^\dagger$.

Other categories such as the \textit{bordism bicategory} of \cite{henriques2013threetier} may be used to introduce additional structure on the input side of the functor, thus defining other types of QFTs such as extended TQFTs \cite{Baez_1995} or CFTs \cite{Runkel:2005qw, henriques2013threetier}.

%%%%%%%%%%%%%%%%%%%%%%%%%%%%%%%
\section{Classification of 2D Frobenius Algebras}\label{app:frobenius}
%%%%%%%%%%%%%%%%%%%%%%%%%%%%%%%

2D TQFT with $\dim\mathcal{H}_{S^1} = 2$ ($\mathcal{H}\cong \mathbb{C}^2$) is the case most analogous to quantum circuits operating on qubits. This is the simplest case in which a notion of complexity exists.\footnote{E.g., an invertible TQFT---one whose Hilbert space on any spatial manifold is spanned by a single state and that is nonzero on all bordisms---has no notion of complexity.}

2D Frobenius algebras can be classified completely up to isomorphism \cite{Fenyes:2015frb}. Since all 2D Frobenius algebras are automatically commutative, such a classification amounts to a classification of 2D TQFTs with $\dim\mathcal{H}_{S^1} = 2$. To state the results of \cite{Fenyes:2015frb}, we fix an orthonormal basis $\{\ket{0}, \ket{1}\}$ for $\mathcal{H}_{S^1}$. We denote the unit, counit, multiplication, comultiplication, and Frobenius form by $\eta$, $\varepsilon$, $\mu$, $\delta$, and $\sigma$, respectively, and we set $\eta(1)\equiv \ket{0}$ by convention. (The canonical reference state is the image of $1\in \mathbb{C}$ under the cup, which is the multiplicative identity, but we are free to choose a basis.) Then 2D Frobenius algebras fall into two classes, depending on the Frobenius form:
\begin{itemize}
\item[(I)] Every 2D TQFT with $\dim\mathcal{H}_{S^1} = 2$ and $\sigma(\ket{00}) = 0$ takes the form
\begin{equation}
\varepsilon : \begin{cases}
\ket{0} \mapsto 0,\\
\ket{1} \mapsto 1,
\end{cases}\quad
\mu : \begin{cases}
\ket{00} \mapsto \ket{0},\\
\ket{01} \mapsto \ket{1},\\
\ket{10} \mapsto \ket{1},\\
\ket{11} \mapsto c\ket{0},
\end{cases}\quad
\delta : \begin{cases}
\ket{0} \mapsto \ket{01} + \ket{10},\\
\ket{1} \mapsto c\ket{00} + \ket{11},
\end{cases}
\label{sigmazero}
\end{equation}
for some $c\in \mathbb{C}$.
\item[(II)] Every 2D TQFT with $\dim\mathcal{H}_{S^1} = 2$ and $\sigma(\ket{00}) \neq 0$ takes the form
\begin{equation}
\varepsilon : \begin{cases}
\ket{0} \mapsto d,\\
\ket{1} \mapsto 0,
\end{cases}\quad
\mu : \begin{cases}
\ket{00} \mapsto \ket{0},\\
\ket{01} \mapsto \ket{1},\\
\ket{10} \mapsto \ket{1},\\
\ket{11} \mapsto \ket{0} + c\ket{1},
\end{cases}
\delta : \begin{cases}
\ket{0} \mapsto \dfrac{1}{d}(\ket{00} + \ket{11}),\vspace{0.2cm}\\
\ket{1} \mapsto \dfrac{1}{d}(\ket{01} + \ket{10} + c\ket{11}),
\end{cases}
\label{sigmanonzero}
\end{equation}
for some $c, d\in \mathbb{C}$ with $d\neq 0$.
\end{itemize}
One can also classify 2D $H^\ast$-algebras. An $H^\ast$-algebra is a Frobenius algebra for which the Frobenius form defines an inner product with respect to which $\delta = \mu^\dag$ and $\varepsilon = \eta^\dag$ (the inner product on $\mathcal{H}_{S^1}^{\otimes N}$ is defined in the canonical way in terms of that on $\mathcal{H}_{S^1}$). The classification implies:
\begin{itemize}
\item[(II$^*$)] Every unitary 2D TQFT with $\dim\mathcal{H}_{S^1} = 2$ takes the form \eqref{sigmanonzero} with $c, d\in \mathbb{R}$, $d > 0$, and
\begin{equation}
\sigma(\ket{00}) = \sigma(\ket{11}) = d, \qquad \sigma(\ket{01}) = \sigma(\ket{10}) = 0.
\label{unitary}
\end{equation}
\end{itemize}
For illustration, note that the $n = 2$ cases of the examples discussed in Section \ref{subsubsec:nilpotentEx} and \ref{subsubsec:semisimpleEx} fit into the above classification. The Frobenius algebra based on $\mathbb{C}[x]/(x^2)$ is a 2D TQFT of type I with $c = 0$, while the Frobenius algebra based on $\mathbb{C}[x]/(x^2 - 1)$ is a 2D TQFT of type II$^*$ with $c = 0$ and $d = 1$.

\renewcommand{\baselinestretch}{1.033} % 1.1
\bibliographystyle{jhep}
\bibliography{TPCbib}
\end{document}